\documentclass[12pt]{iopart}
\usepackage[backend=biber,style=phys,biblabel=brackets,,maxbibnames=4]{biblatex}
\usepackage{subcaption}
\usepackage{graphicx}
\newcommand{\solarmass}{M\textsubscript{\(\odot\)}}
\begin{document}
\title[GSpyNetTree]{GSpyNetTree: A signal-vs-glitch classifier for gravitational-wave event candidates }

\author{Sof\'ia \'Alvarez-L\'opez$^{1,2,3}$, Annudesh Liyanage$^1$, Julian Ding$^{1,4}$, Raymond Ng$^4$ and Jess McIver$^1$}
\address{$^1$ Department of Physics and Astronomy, University of British Columbia, Vancouver, British Columbia, V6T1Z4, Canada}
\address{$^2$ Departamento de F\'isica, Universidad de los Andes, 111711, Bogot\'a, Colombia}
\address{$^3$ Departamento de Ingenier\'ia de Sistemas y Computaci\'on, Universidad de los Andes, 111711, Bogot\'a, Colombia}
\address{$^4$ Department of Computer Science, University of British Columbia, Vancouver, British Columbia, V6T1Z4, Canada}
\ead{sofia.alvarez@ligo.org} 
\vspace{10pt}
\begin{indented}
\item[]April 2023
\end{indented}

\begin{abstract}
Despite achieving sensitivities capable of detecting the extremely small amplitude of gravitational waves (GWs), LIGO and Virgo detector data contain frequent bursts of non-Gaussian transient noise, commonly known as `glitches'. Glitches come in various time-frequency morphologies, and they are particularly challenging when they mimic the form of real GWs. Given the higher expected event rate in the next observing run (O4), LIGO-Virgo GW event candidate validation will require increased levels of automation. Gravity Spy, a machine learning tool that successfully classified common types of LIGO and Virgo glitches in previous observing runs, has the potential to be restructured as a signal-vs-glitch classifier to accurately distinguish between glitches and GW signals. A signal-vs-glitch classifier used for automation must be robust and compatible with a broad array of background noise, new sources of glitches, and the likely occurrence of overlapping glitches and GWs. We present GSpyNetTree, the Gravity Spy Convolutional Neural Network Decision Tree: a multi-CNN classifier using CNNs in a decision tree sorted via total GW candidate mass tested under these realistic O4-era scenarios.

\end{abstract}

\vspace{2pc}

\vspace{1pc}

\noindent \submitto{\CQG}

\section{Introduction}

Since the first observing run (O1) \cite{1stdiscovery, o1era_detectors}, and following major upgrades \cite{o1calib, o2detector, o3detector}, the Advanced LIGO (aLIGO) \cite{aLIGO} and Advanced Virgo (AdVirgo) \cite{AdVirgo} detectors have made dozens of transient GW signal detections in the second (O2) \cite{o2} and third (O3a, O3b) \cite{o3a, gwtc21, o3b} observing runs. Due to the extreme and increasing sensitivity of these detectors, they are prone to non-astrophysical noise sources (glitches) that can mask or mimic true GW signals \cite{1stnoise}. These glitches are particularly problematic as they may generate false-positive candidates \cite{massinger, derek, derek_validation}, corrupt data, and bias astrophysical parameter estimation \cite{powell_jade, macas, curious_case}. Along with the vast amount of data that LIGO and Virgo generate, developing robust tools and methods to automatically identify and characterize these glitches is crucial for extracting GW signals from the data.

A common tool used for glitch classification is Gravity Spy \cite{gspy, gspyO3}: a Convolutional Neural Network (CNN) image classifier that distinguishes 21 different glitch classes and 1 Chirp class (consisting of data from hardware injections that emulate the behavior of GWs by displacing the detector's test masses \cite{hwinjections}) via time-frequency visualizations, a type of spectrograms called omegascans \cite{chatterji}. A CNN is a type of Deep Learning algorithm designed for image classification consisting of an input layer, an output layer, and several hidden layers that extract useful features from the inputs.

After successfully classifying glitches in previous observing runs \cite{jane}, Gravity Spy's architecture has shown the potential to be used as a signal-vs-glitch classifier. This new application is particularly relevant as detectors become more sensitive for the next observing run (O4), and further automation of event candidate validation is required \cite{derek_validation}.

Numerous investigations using other Machine Learning techniques in glitch classification, such as Principal Component Analysis (PCA) and clustering glitches with Deep Transfer Learning, have been explored \cite{powell, uiuc}. Other approaches also include iDQ, a supervised learning tool that detects detector noise artifacts through auxiliary channel data \cite{idq}, and GWSkyNet, a low-latency classifier for public GW candidates that requires information from multiple detectors \cite{gwskynet}. In order to fully leverage CNNs as the state-of-the-art method for complex image classification tasks \cite{cnn}, we build on Gravity Spy's CNN classification approach to distinguish signals from glitches for future observing runs using single detector strain $h(t)$ data.

This paper builds upon the proof of principle described in Jarov et al. \cite{seraphim}, which outlines a new multi-classifier method to leverage prior Gravity Spy architecture to distinguish GWs from glitches. Considering previous investigations on improving Gravity Spy that have shown that its inaccuracies in glitch classification tend to be higher in poorly represented classes in the CNN's training set \cite{bahaadini}, Jarov et al.'s study recommends significant changes to Gravity Spy for the purpose of a signal-vs-glitch classifier. First, it recommends augmenting the data of the Chirp class, which is morphologically similar to the typical GW events seen in O3 \cite{gspyO3}. However, instead of using hardware injections, it uses GW software simulations that allow the generation of more GW samples, compared to the severely underrepresented Chirp class in the original Gravity Spy training set \cite{gspy}. It also recommends deploying specialized training sets to handle different ranges of total candidate signal mass, as low and high-mass mergers have very distinct morphologies, which may be more prone to confusion with particular glitch classes. 

We developed this proposed signal-vs-glitch multi-classifier architecture and analyzed its readiness for O4, as described in Section \ref{methods}. After generating the specialized training sets based on GW and glitch morphology and incorporating the data augmentation suggested by Jarov et al. \cite{seraphim}, we built three different classifiers, one per training set, with the same CNN architecture Gravity Spy leverages. With the three signal-vs-glitch classifiers, we made a decision tree sorted via total GW candidate mass, constituting the base for GSpyNetTree, the Gravity Spy Convolutional Neural Network Decision Tree. During O4, this tool will intake GW candidate events from GraceDB \cite{gracedb} and classify them as GWs or glitches as part of the LIGO-Virgo Data Quality Report \cite{dqr}.

After building GSpyNetTree according to the recommendations in Jarov et al. \cite{seraphim}, we noted that the original Gravity Spy CNN architecture could be further improved. We decided to use Inception V3 \cite{inceptionv3}, Google's state-of-the-art CNN for image classification tasks, as explained in section \ref{results}. We noted a vast improvement in classification accuracy, particularly for the GW class. We also considered the additional changes expected for O4, including different persistent noise subtraction for calibrated data, new potential noise sources, a high expected detection rate \cite{detectionRate}, and the likely occurrence of overlapping glitches and GW signals in time-frequency visualizations.

Section \ref{valstudies} presents three validation studies based on the classification challenges posed by the O4 sensitivity increase. Additionally, we propose solutions to overcome them for a future O4-era version of GSpyNetTree. First, we study how the current version of GSpyNetTree responds to data with non-linear subtraction of 60 Hz AC power artifacts, as expected for low latency LIGO data in O4. We evaluate its readiness for the new background noise expected in O4 and its transferability to Virgo data, which has a different noise background from the two LIGO detectors. We also test how GSpyNetTree responds to glitches not included in its training set, as new noise sources are expected to appear during O4. We curate a variety of glitches covering several cases of interest, including frequently occurring glitches and glitches morphologically similar to others already included in the training set. We then evaluate how GSpyNetTree responds to GW candidates overlapping with glitches, which is even more likely to happen during O4 than O3, where 24\% of candidates overlapped with one or more glitches \cite{o3a, o3b}. We propose a new multi-label architecture that allows GW classification, even in the case of glitches overlapping with signals. Finally, we comment on prospects for GSpyNetTree in the O4 era.

\begin{figure}[htbp]
\centering
\captionsetup[subfigure]{font=small}
     \begin{subfigure}[b]{0.98\textwidth}
         \includegraphics[width=\textwidth]{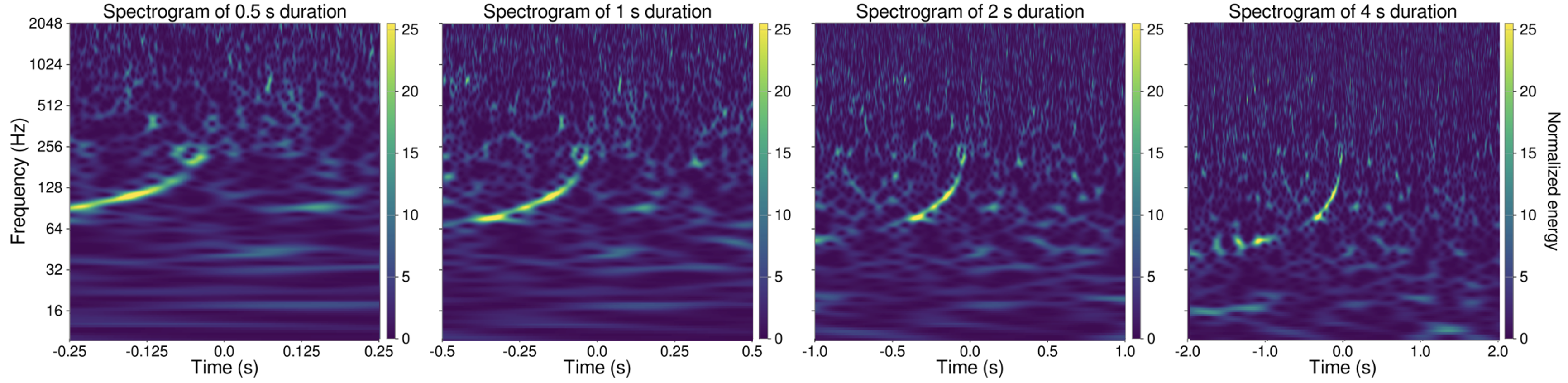}
         \caption[]{} 
         \label{fig:lm}
     \end{subfigure}
     \hfill
     \begin{subfigure}[b]{\textwidth}
         \includegraphics[width=\textwidth]{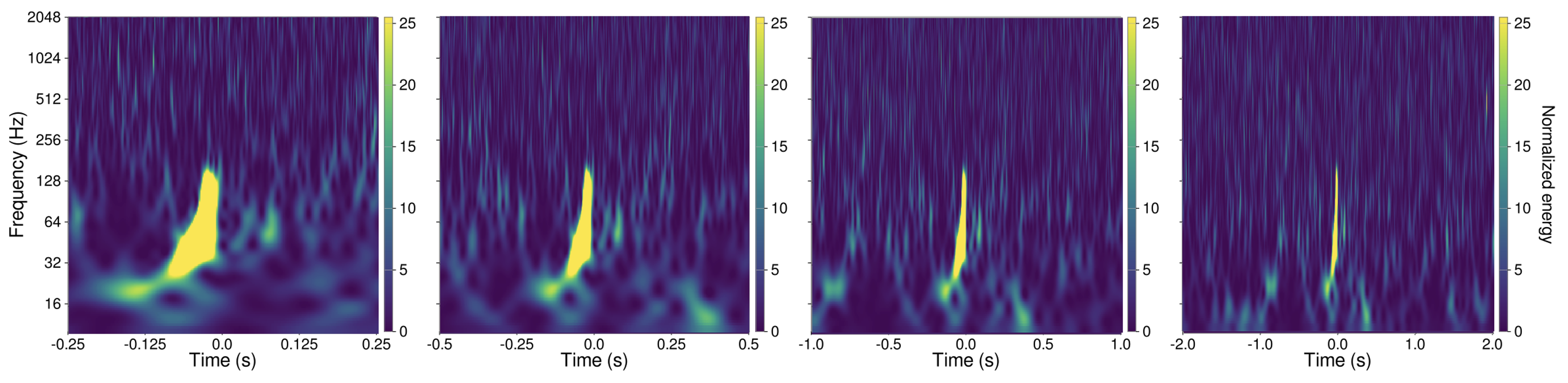}
         \caption[]{} 
         \label{fig:hm_highsnr}
     \end{subfigure}
     \hfill
     \begin{subfigure}[b]{\textwidth}
         \includegraphics[width=\textwidth]{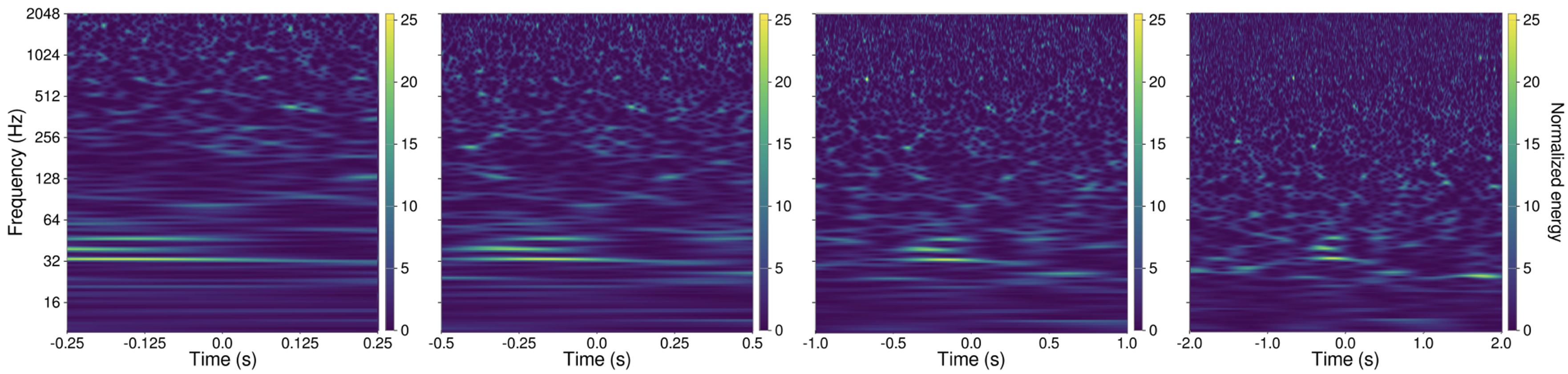}
         \caption[]{ } 
         \label{fig:hm_lowsnr}
     \end{subfigure}
     \hfill
     
      \begin{subfigure}[b]{\textwidth}
         \includegraphics[width=\textwidth]{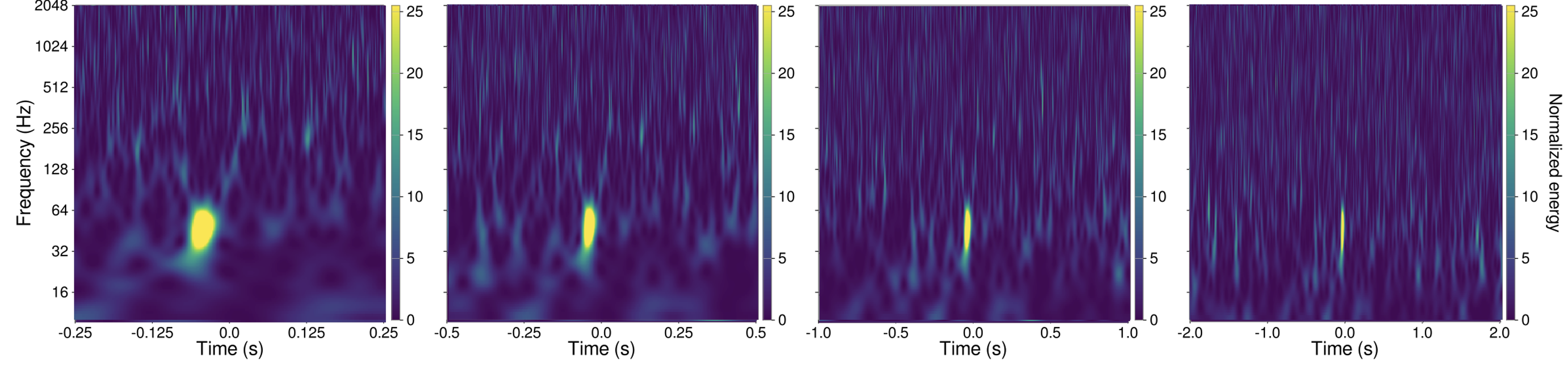}
         \caption[]{} 
         \label{fig:ehm}
     \end{subfigure}
        \caption{\small Examples of spectrograms of simulated GW signals with all four durations used in the training set of GSpyNetTree (0.5 s, 1 s, 2 s, and 4 s). (a) Software simulated GW signal in the low mass regime, with a total mass of 29.2 $\solarmass$. (b) Simulated GW signal in the high mass regime, with a total mass of 118.7 $\solarmass$. (c) Simulated GW signal in the high mass regime, with a total mass of 182 $\solarmass$, but with a significantly lower signal-to-noise ratio (SNR) than the example shown in Figure \ref{fig:hm_highsnr}. Signals with a low SNR may occur in any of the mass ranges, and they may be similar to the No Glitch class (Figure \ref{fig:noglitch}). (d) Simulated GW signal in the extremely high mass regime, with a total mass of 283 $\solarmass$. These signals are morphologically similar to Low-frequency blips (Figure \ref{fig:lfblip}).     } 
        \label{fig:gw_samples}
\end{figure}

\section{Building GSpyNetTree and its training datasets}\label{methods}

\begin{figure}[htbp]
\centering
\captionsetup[subfigure]{font=small}
     \begin{subfigure}[b]{0.95\textwidth}
         \includegraphics[width=\textwidth]{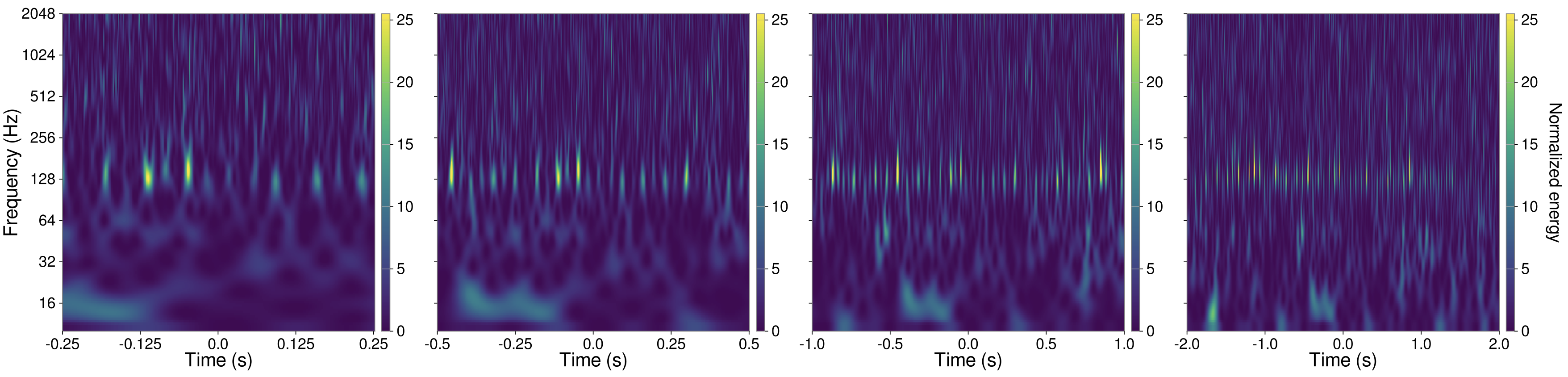}
         \caption[]{ Example of a Scratchy glitch.}
         \label{fig:blip}
     \end{subfigure}
     \begin{subfigure}[b]{0.95\textwidth}
         \includegraphics[width=\textwidth]{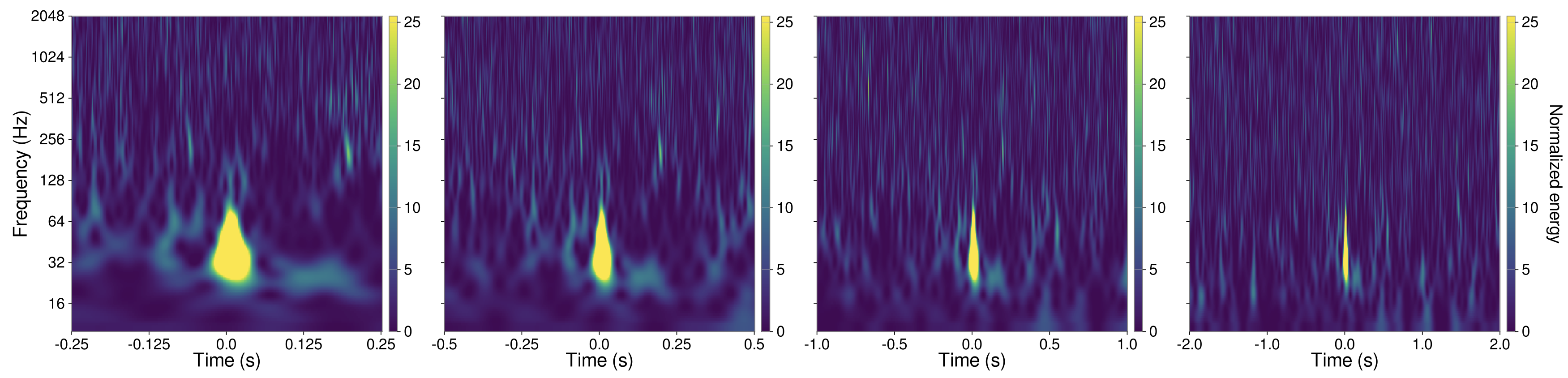}
         \caption[]{Example of a Low-frequency blip glitch.}
         \label{fig:lfblip}
     \end{subfigure}
     \begin{subfigure}[b]{0.95\textwidth}
         \includegraphics[width=\textwidth]{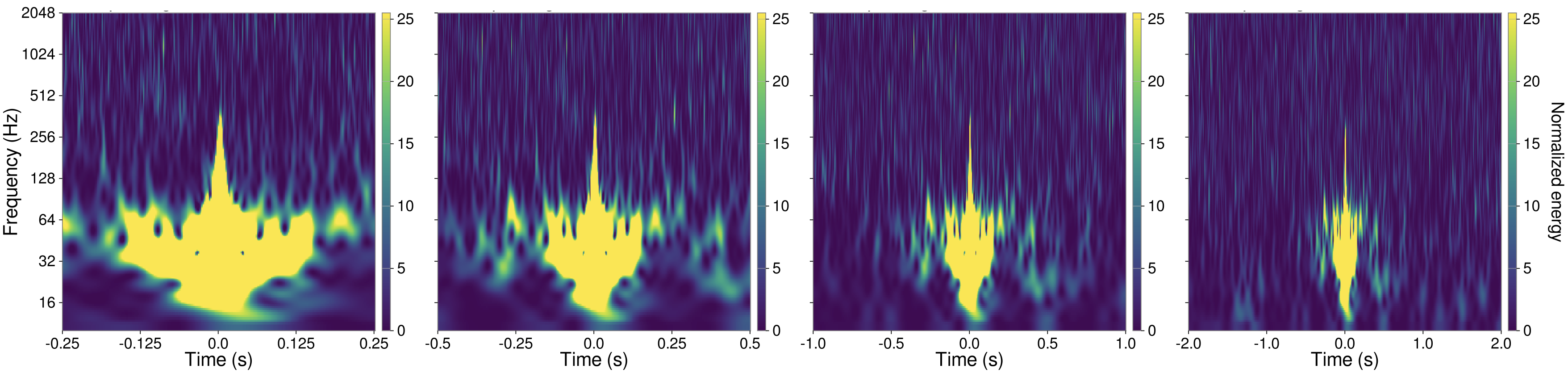}
         \caption[]{Example of a Koi Fish glitch.}
         \label{fig:koi}
     \end{subfigure}
      \begin{subfigure}[b]{0.95\textwidth}
         \includegraphics[width=\textwidth]{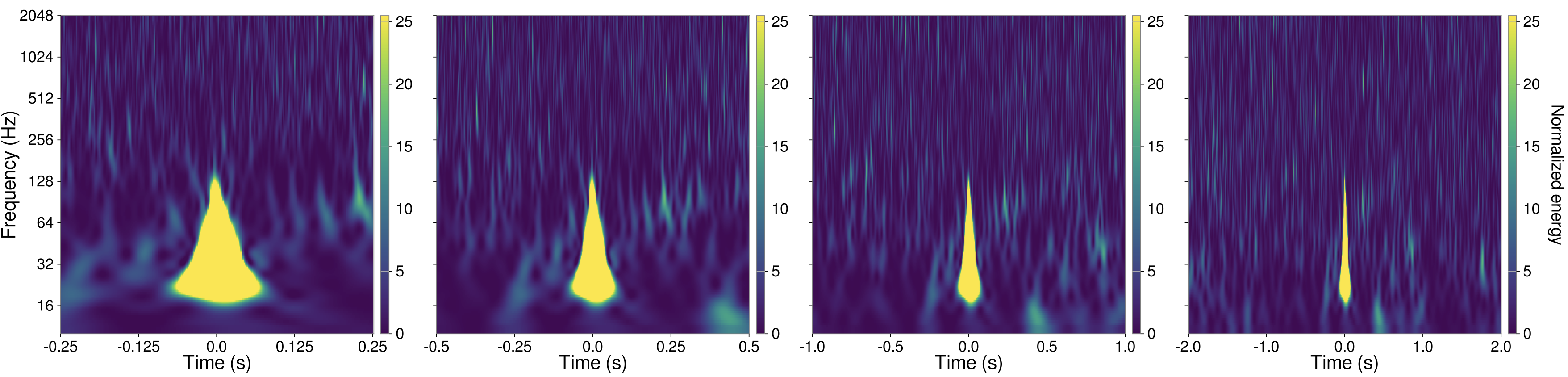}
         \caption[]{Example of a Tomte glitch.}
         \label{fig:tomte}
     \end{subfigure}
      \begin{subfigure}[b]{0.95\textwidth}
         \includegraphics[width=\textwidth]{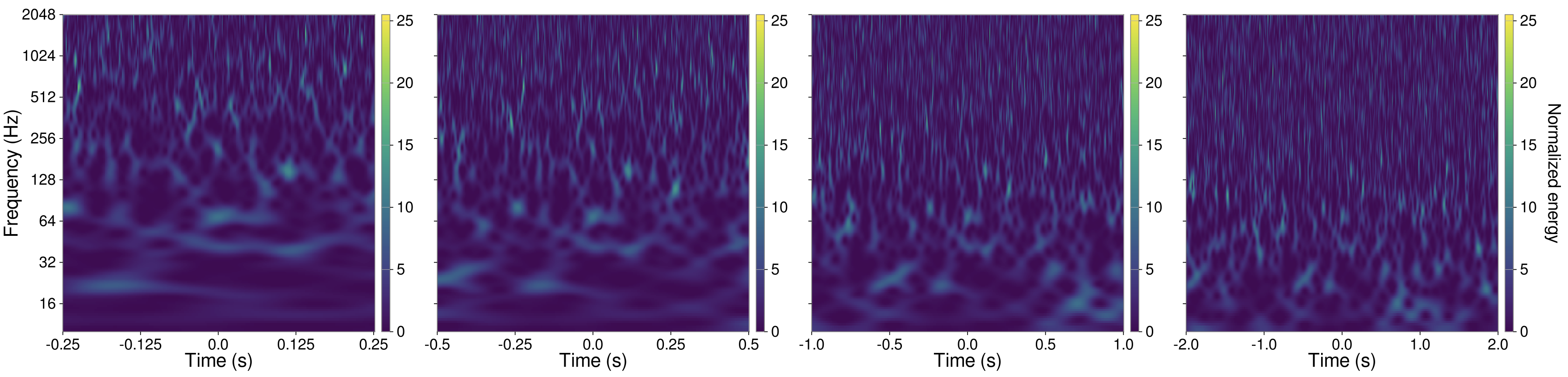}
         \caption[]{Example of a No Glitch.} 
         \label{fig:noglitch}
     \end{subfigure}
        \caption{Examples of selected glitches with all four durations used in GSpyNetTree (0.5 s, 1 s, 2 s, and 4 s). These non-astrophysical events are morphologically similar to the GW signals in the three mass ranges. All samples fetched from LIGO-DV web \cite{ligodv}. } 
        \label{fig:glitch_samples}
\end{figure}

Building upon the proof of principle described in Jarov et al. \cite{seraphim}, GSpyNetTree leverages a decision tree of three CNN classifiers, each trained on a specialized and balanced set of GWs and morphologically similar glitches, sorted via estimated candidate mass metadata. The three classifiers are: the low-mass (LM) classifier (for candidates with an estimated total mass below 50 {\solarmass}), the high-mass (HM) classifier (for candidates with an estimated total mass between 50 {\solarmass} and 250 {\solarmass}), and the extremely high-mass (EHM) classifier (for candidates with an estimated total mass above 250 {\solarmass}). Depending on the mass estimate provided via GraceDB \cite{gracedb}, each candidate GW event is sent to the LM CNN or HM CNN to determine whether it is astrophysical or a glitch. If the candidate's mass is above 250 $\solarmass$, it is then sent from the HM classifier to the EHM classifier for more accurate classification. Figure \ref{fig:gw_samples} shows examples of simulated GW signals in each of the mass ranges, and Figure \ref{fig:glitch_samples} shows the glitches they are morphologically similar to. Table \ref{tab:class_distribution} specifies the glitch classes considered for each mass range depending on morphological similarities, as considered by Jarov et al. \cite{seraphim}. 

\begin{table}[htbp]
\resizebox{\textwidth}{!}{%
\begin{tabular}{|cc|cc|cc|}
\hline
\multicolumn{2}{|c|}{\textbf{Low-Mass (LM) Classifier}}        & \multicolumn{2}{c|}{\textbf{High-Mass (HM) Classifier}}         & \multicolumn{2}{c|}{\textbf{Extremely High-Mass (EHM) Classifier}} \\ \hline
\multicolumn{1}{|c|}{\textbf{Class}}         & \textbf{Samples} & \multicolumn{1}{c|}{\textbf{Class}}           & \textbf{Samples} & \multicolumn{1}{c|}{\textbf{Class}}             & \textbf{Samples}  \\ \hline
\multicolumn{1}{|c|}{GW (3-50 {\solarmass})} & 1000            & \multicolumn{1}{c|}{GW (50-250 {\solarmass})} & 1000            & \multicolumn{1}{c|}{GW (250-350 {\solarmass})}  & 1000             \\
\multicolumn{1}{|c|}{Blip}                   & 999             & \multicolumn{1}{c|}{Blip}                     & 999             & \multicolumn{1}{c|}{Blip}                       & 999              \\
\multicolumn{1}{|c|}{Low-Frequency Blip}     & 1039            & \multicolumn{1}{c|}{Low-Frequency Blip}       & 1039            & \multicolumn{1}{c|}{Low-Frequency Blip}         & 1039             \\
\multicolumn{1}{|c|}{No Glitch}              & 1017            & \multicolumn{1}{c|}{No Glitch}                & 1017            & \multicolumn{1}{c|}{No Glitch}                  & 1017             \\
\multicolumn{1}{|c|}{Scratchy}               & 1093            & \multicolumn{1}{c|}{Koi Fish}                 & 990             & \multicolumn{1}{c|}{}                           &                  \\
\multicolumn{1}{|c|}{}                       &                 & \multicolumn{1}{c|}{Tomte}                    & 758             & \multicolumn{1}{c|}{}                           &                  \\ \hline
\end{tabular}%
}
\caption{Classes and number of samples (before time-offset augmentation, which add four more examples for each sample listed) per class for each of the GSpyNetTree’s classifiers. The GW mass ranges listed indicate the target total mass of an event candidate.}
\label{tab:class_distribution}
\end{table}

The EHM classifier presents additional challenges that need to be addressed. Low-frequency blips share strong similarities in duration, frequency range, and morphology with EHM mergers. 
Indeed, the original Gravity Spy model misclassifies EHM mergers as Low-frequency blips with 99\% confidence \cite{seraphim}. However, since the mass range of detected GWs is expected to increase in each observing run, it is essential to make GSpyNetTree robust for these possible future detections. GSpyNetTree incorporates a spectrogram scaling technique that Jarov et al. showed to be of great utility in this mass range \cite{seraphim}: we apply the Mercator projection, which stretches the signals vertically, scaling the image features to better segregate GW signals from Low-frequency blips. 

Additionally, when dealing with CNN training sets, it is important to consider data augmentation techniques to avoid overfitting. One strategy is to generate slightly modified versions of existing samples, increasing the amount of training data. Thus, we generated four random time offsets within $0.1\; \mathrm{s}$ in the time-frequency visualizations of each GW and glitch so that samples are not always perfectly centered. This also makes GSpyNetTree robust to small offsets in estimated candidate merger times.

Following this approach, we built an augmented training set for each of GSpyNetTree's CNNs. First, we ensured a balanced representation of both GWs and glitches, as previous studies have shown that a higher rate of inaccuracies is related to poorly represented classes in Gravity Spy \cite{bahaadini, seraphim}. Additionally, it is well known for CNN image classifiers that increasing the size of the training set improves classification performance ~\cite{dataset_size_performance}. Instead of having $\sim 150$ instances per class as in previous studies \cite{seraphim}, we decided to enrich our training sets with more samples so that there were $1000 \pm 300$ of each per class. Moreover, we included the No Glitch class for the three classifiers to account for GWs with low signal-to-noise ratio (SNR). Table \ref{tab:class_distribution} shows the distribution of classes per classifier in the GSpyNetTree training sets.

We fetched all the glitches included in this training set from Gravity Spy classifications via LIGO-DV web \cite{ligodv} for both LIGO Hanford (LHO) and LIGO Livingston (LLO) observatories. Additionally, we manually verified all examples to discard misclassified or morphologically unconventional samples, which we saved for the validation study explained in section \ref{exotic}. For the GW simulations, we identified several segments of 64-second quiet detector data for both LHO and LLO during previous observing runs and injected simulated waveforms into them using the inspiral injection module of LALSuite \cite{lalsuite}, using the waveform model \texttt{IMRPhenomPv2} \cite{waveform1, waveform2}. GSpyNetTree's GW examples are uniformly drawn from a total merger mass range of $5 {\solarmass}$ to $350 {\solarmass}$, with individual masses ranging from $2 {\solarmass}$ to $175 {\solarmass}$, an SNR range of 8 to 35, and individual component spins ranging from 0.05 to 0.95. 

\begin{figure}[htbp]
    \centering
    \includegraphics[width=0.9\textwidth]{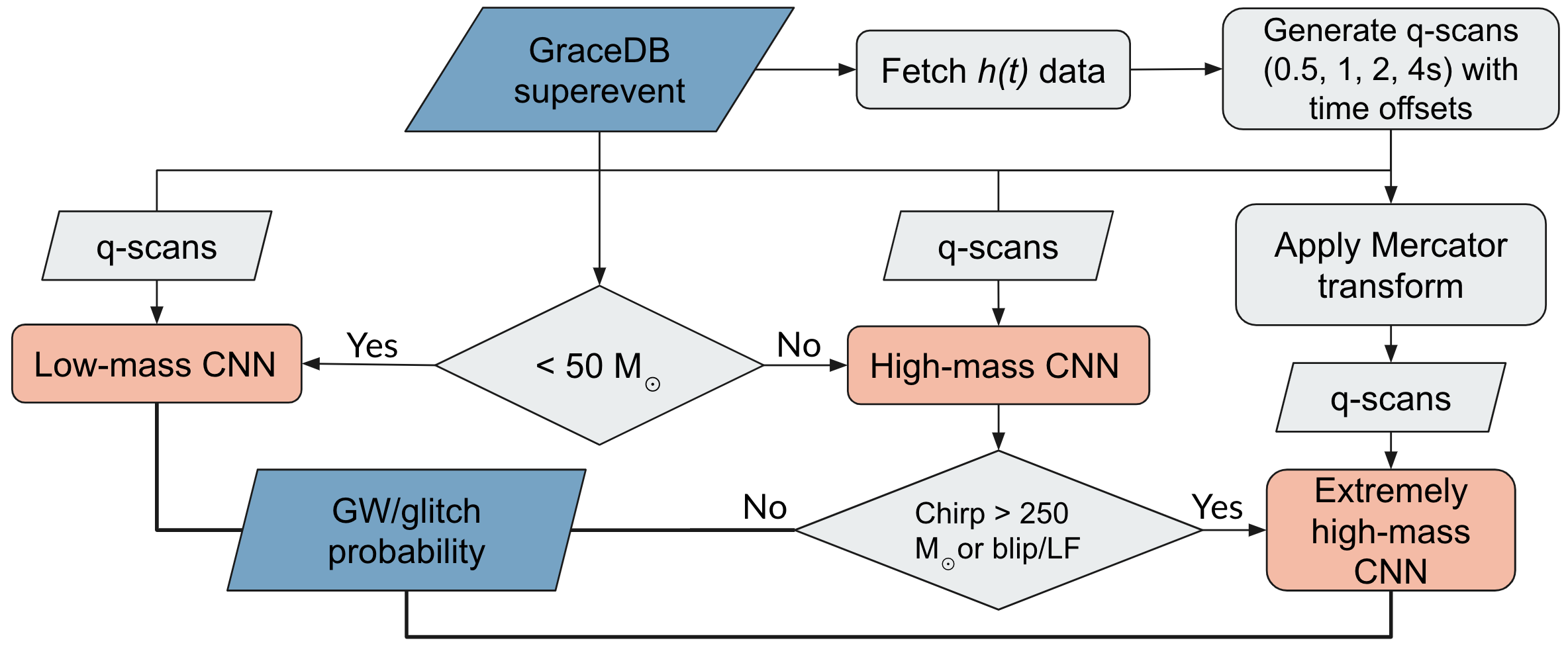}
    \caption{GSpyNetTree architecture: Triggered by a GraceDB superevent \cite{gracedb}, time series (strain) data is fetched to generate spectrograms of 0.5, 1, 2, and 4 second durations. Time-frequency spectrogram visualizations are sent to the classifiers based on the estimated candidate merger mass, and the Mercator transform is applied to the extremely high mass GW candidate visualizations. Each CNN outputs the probability that the input visualization contains a GW, an included class of glitch, or no glitch.}
    \label{fig:gspynettree}
\end{figure}

We used the strain $h(t)$ time series from each glitch and simulated GW to generate the time-frequency features required by GSpyNetTree for each sample: four spectrograms 0.5 s, 1 s, 2 s, and 4 s in duration arranged in a $2\times2$ matrix, as in the original Gravity Spy architecture \cite{gspy}. The different spectrogram durations are used to better capture GW mergers and glitches of different durations. We fed these samples to GSpyNetTree, with architecture shown in Figure \ref{fig:gspynettree}. After applying the time-offset augmentation, each sample is directed to one of GSpyNetTree's CNNs, based on its estimated total mass. If the sample's total mass is less than 50 $\solarmass$, it is directed to the LM classifier; otherwise, it is sent to the HM classifier. Events classified by the HM classifier are further directed to the EHM classifier when the following criteria are met: the total mass of the candidate is estimated to be greater than 250 $\solarmass$ and the HM classifier has classified the candidate  as a Low-frequency blip, No Glitch, or GW.
In the EHM classifier, the Mercator projection is applied to the sample before classification. Finally, each CNN returns an array of probabilities assigned to each class per sample. Once in production, GSpyNetTree will intake GW candidate events uploaded to GraceDB \cite{gracedb} via the Data Quality Report \cite{dqr} and classify them as GWs or glitches with a reported probability.

Once GSpyNetTree’s training sets were complete, we trained the CNNs and evaluated their results. We used 80\% of the dataset for training,  and allocated the remaining 20\% for testing the CNNs. Within the training set, we allocated 20\% for validation. Section 3 presents these results.

\section{A New Architecture for GSpyNetTree}\label{results}
We first tested GSpyNetTree with the augmented training sets described in Section \ref{methods}, using the original Gravity Spy architecture \cite{gspy}. 

Our low-mass CNN had an overall accuracy of 94\%, the high-mass CNN achieved 94.6\%, while the extremely high-mass CNN made 96\% accurate predictions. Additionally, all of them had a 92\% accuracy for the GW class, with 7.2\%, 4.5\%, and 2.9\% of GW signals misclassified as No Glitches in the LM, HM, and EHM classifiers, respectively. This was expected for low SNR signals, which may appear faint in spectrogram visualizations. Therefore, these misclassifications are not problematic for GSpyNetTree's purposes. While these are good results, it is not desirable to misclassify 8\% of the astrophysical data per mass range, especially considering the high detection rate expected for O4 \cite{detectionRate}.

\begin{figure}[htbp]
    \centering
    \includegraphics[width=\textwidth]{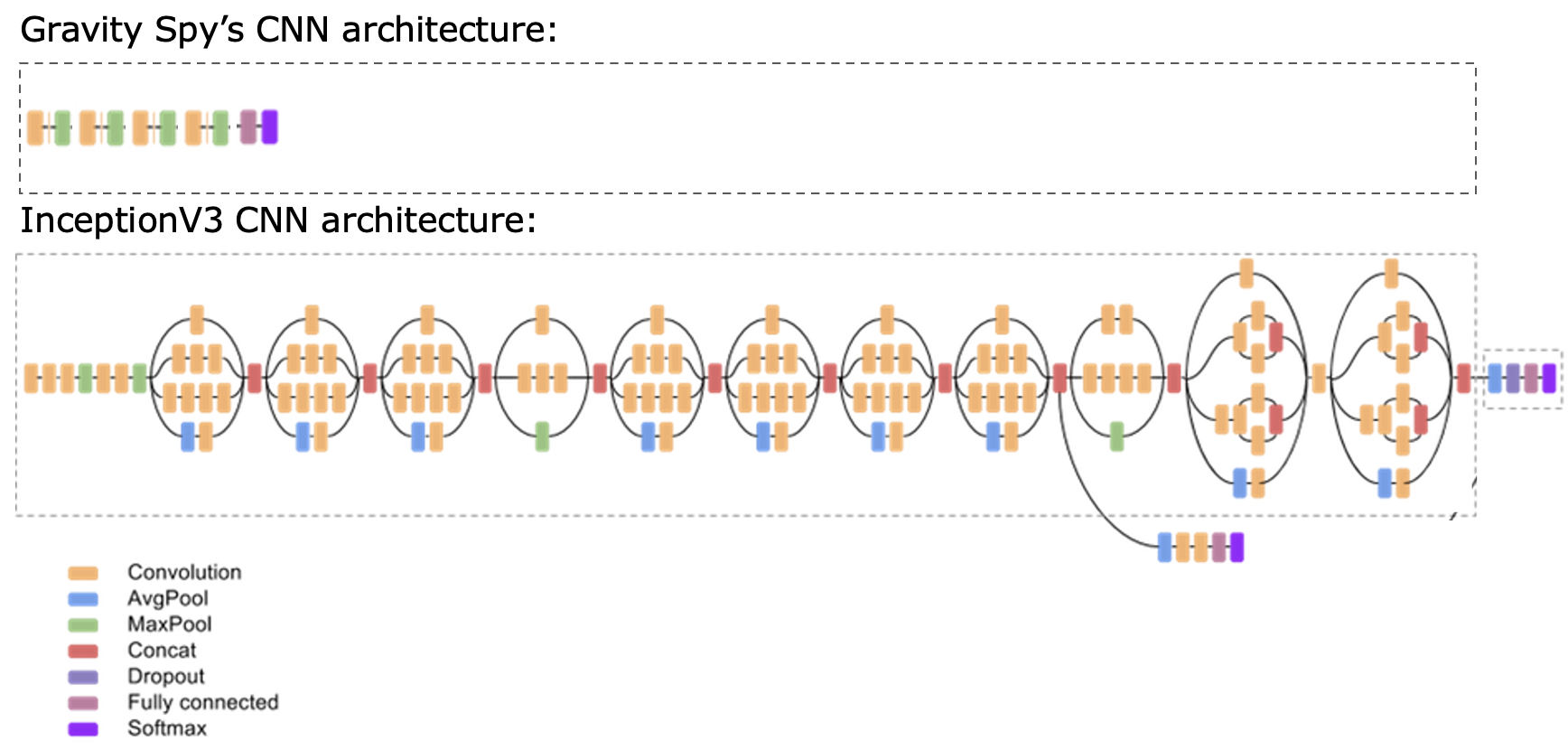}
    \caption{Comparison of the Convolutional Neural Network architecture of Gravity Spy (upper panel) \cite{gspy} and Inception V3’s architecture (lower panel) \cite{inceptionv3}. Inception is a more robust, deeper network, which makes it ideal for complex image classification tasks. (Adapted from \cite{inceptionv3, deeper}).}
    \label{fig:architecture}
\end{figure}

In order to further improve the accuracy of classifications, especially of the GW class, we used a new CNN architecture. It is well known that CNNs are the state-of-the-art method for complex image classification tasks \cite{cnn}. Therefore, several networks specifically designed to tackle these problems have been studied and developed in the computer science realm. One of them is Inception V3: Google's state-of-the-art CNN \cite{inceptionv3}. It is made up of 42 layers, which makes it a very deep model compared to Gravity Spy's 5 layers, as shown in Figure \ref{fig:architecture}. Additionally, it has shown better accuracy, less computational cost, and a very low error (the percentage of erroneously classified samples) in various image classification tasks. Deeper neural networks require larger training sets to avoid overfitting; however, due to our drastically increased training set size over the one used in Jarov et al. \cite{seraphim}, Inception V3 remained a viable option for our investigation.

\begin{figure}[htbp]
    \centering
    \includegraphics[width=0.75\textwidth]{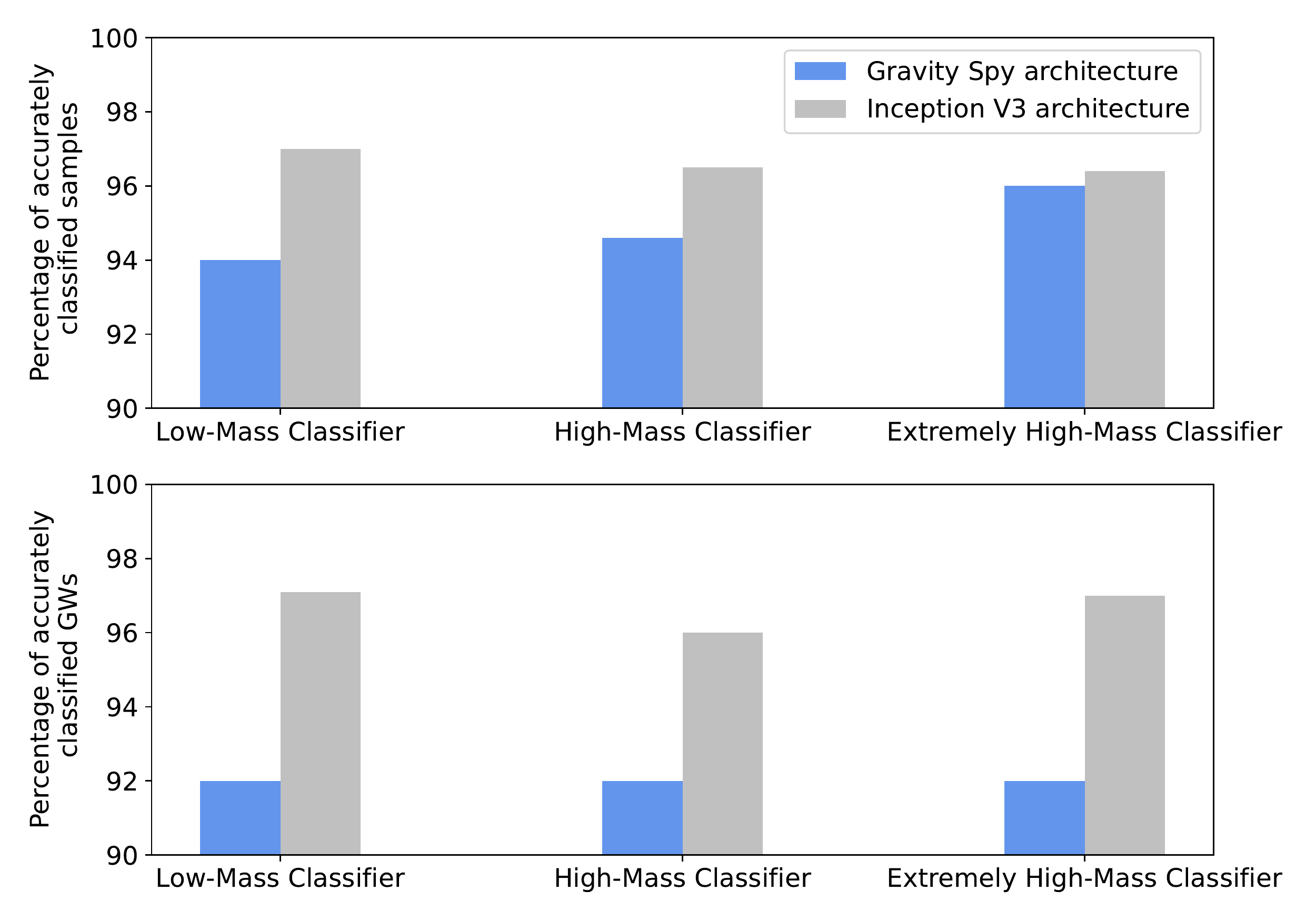}
    \caption{Accuracy results for the three GSpyNetTree CNNs trained using the Gravity Spy architecture (in blue) and the InceptionV3 architecture (in gray) for all GWs and glitches in the test set (upper panel) and for only GWs (lower panel). In the case of chirps, the classification accuracy improved from 92\% in all three cases to 97\% (LM), 96\% (HM), and 97\% (EHM). Additionally, overall accuracy also improved from 94\%, 95\%, and 96\% to 97\%, for the LM, HM, and EHM classifiers respectively.}
    \label{fig:results}
\end{figure}

We trained the three Inception V3 CNNs (LM, HM, and EHM) from scratch, as we had a large enough training set to do so, using the training data introduced in section \ref{methods}. Figure \ref{fig:results} shows the improvement in classification with this new approach. Following the implementation of the Inception V3 CNNs, all classifiers reached more than 96\% accuracy for GWs and all glitch classes, with 3.9\%, 2.6\%, and 2.3\% of GWs misclassified as No Glitches, such that the GW (+ No Glitch) accuracy was 96\% (+3.9\%),  96\% (+2.6\%), and 97\% (+2.3\%) for the LM, HM, and EHM classifiers. This is a considerable decrease in the amount of misclassified astrophysical events.

Once we improved the classification accuracy of GW signals, we validated GSpyNetTree's readiness for O4 by performing three validation studies. We tested both the original Gravity Spy and the Inception V3 architecture, and our results were better with the latter for all validation studies. We present these results in Section \ref{valstudies}.

\section{Validating GSpyNetTree's readiness for O4}\label{valstudies}

\subsection{Testing the CNNs reliance on background}
The first validation study we performed aimed to evaluate the dependence of GSpyNetTree on detector background noise. This is important for O4 because the noise subtraction used to produce low-latency calibrated $h(t)$ is expected to differ from previous observing runs. O4 low-latency data is expected to use a non-linear subtraction of AC 60 Hz power artifacts, similar to the technique used for publically released data from the O3 run \cite{gwosc}. It is also important to evaluate how transferable the GSpyNetTree model is to detect GWs for Virgo and KAGRA detectors, which have a different noise background from LIGO. 

\begin{figure}[htbp]
  \centering
    \includegraphics[scale=0.42]{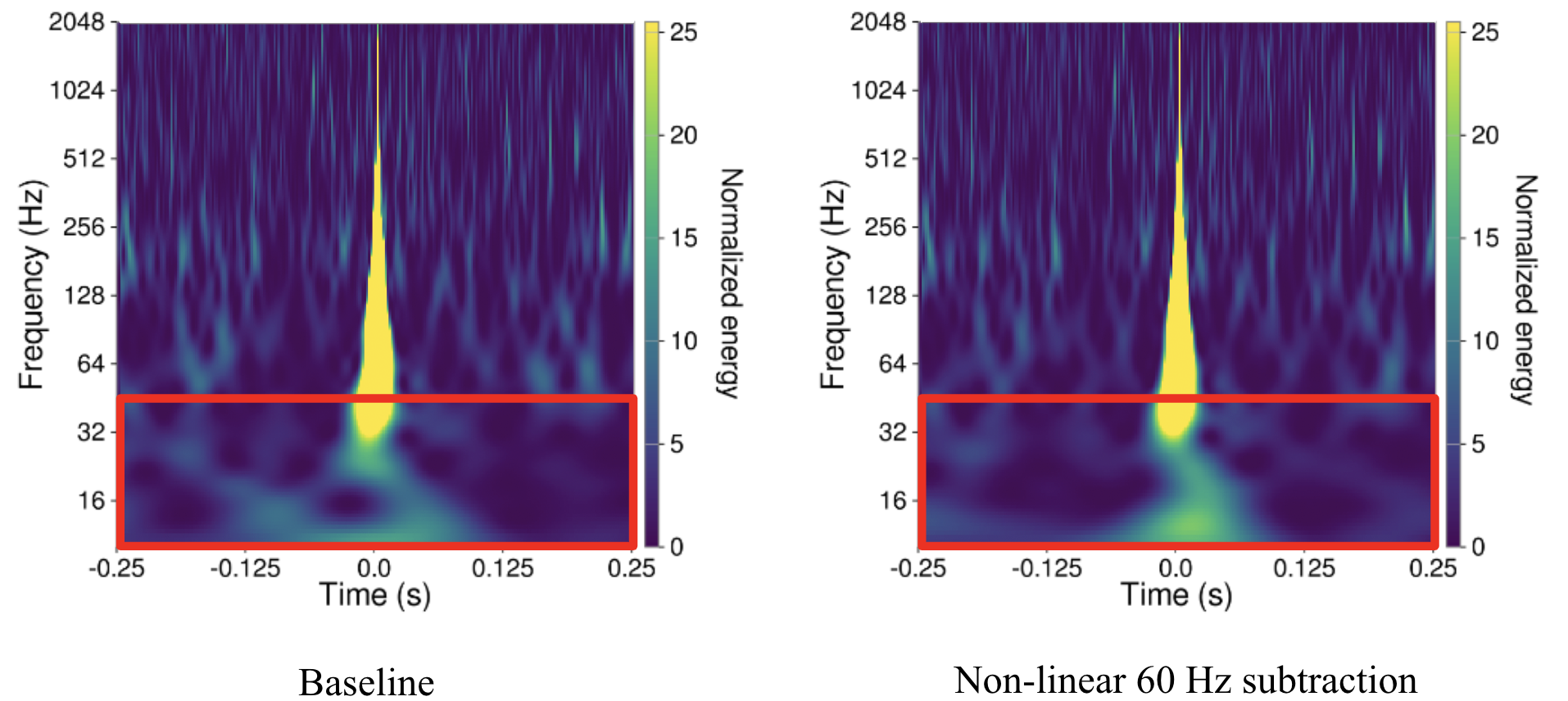}
    \caption{Spectrograms of an O3-era Blip from the Hanford detector, fetched from the original low-latency strain channel (left) and the higher-latency channel with a non-linear subtraction applied (right). A subtle difference in background can be seen around 60 Hz, as shown in the red boxes. The non-linear subtraction is expected to be implemented for low-latency data in O4. }
    \label{fig:denoised}
\end{figure}

To test the CNNs’ reliance on background, we used 100 glitch examples per glitch class, using the non-linear 60 Hz subtracted strain channel, for each classifier. Figure \ref{fig:denoised} shows an example of a Blip fetched from the original (left) and the clean (right) channels. 

\begin{figure}[htbp]
  \centering
    \includegraphics[scale=0.45]{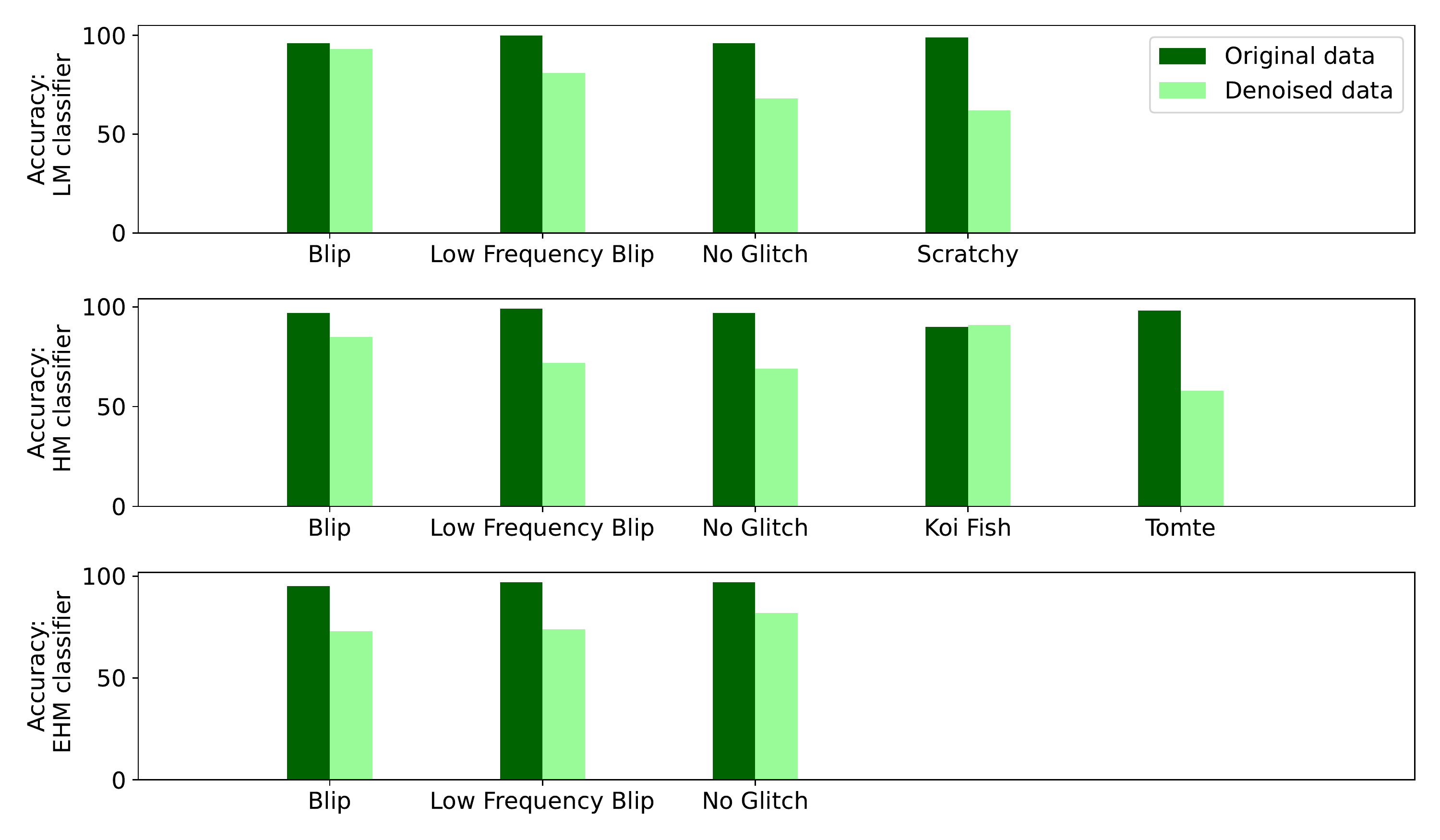}
    \caption{Accuracy results for the LM (top), HM (middle), and EHM (bottom) classifiers for each of the glitches in the validation set. Except for Koi Fish glitches, there is a significant decrease ($\sim 20\%$) in accuracy when denoised background data is used. }
    \label{fig:denoised_barchart}
\end{figure}

Even though both spectrograms look quite similar to the naked eye, these small differences in background significantly impact CNN accuracy. In fact, when tested with non-linearly noise subtracted data, overall accuracy decreased to 76\%, 75\%, and 76\% for the LM, HM, and EHM classifiers, respectively. Additionally, as shown in Figure \ref{fig:denoised_barchart}, individual class accuracy also decreased for all of the glitches but one (Koi Fish). This likely occurs because, as shown in Figure \ref{fig:koi}, these glitches usually cover a wide frequency and time range, so the denoised background data does not significantly impact the visualization.

Following this discovery, we note that to be able to transfer GSpyNetTree to O4 with maximum accuracy, the CNN's dependence on background should be addressed with a new augmented training set. Additionally, as only LHO and LLO glitches and background are currently used, we should include Virgo GWs and glitches in our training sets for the O4-era GSpyNetTree. This way, the glitch dataset will be robust enough to account for changes in detector background noise.

\subsection{Testing CNN classification of glitches not included in the original training set}\label{exotic}

\begin{figure}[htbp]
  \centering
    \includegraphics[scale=0.5]{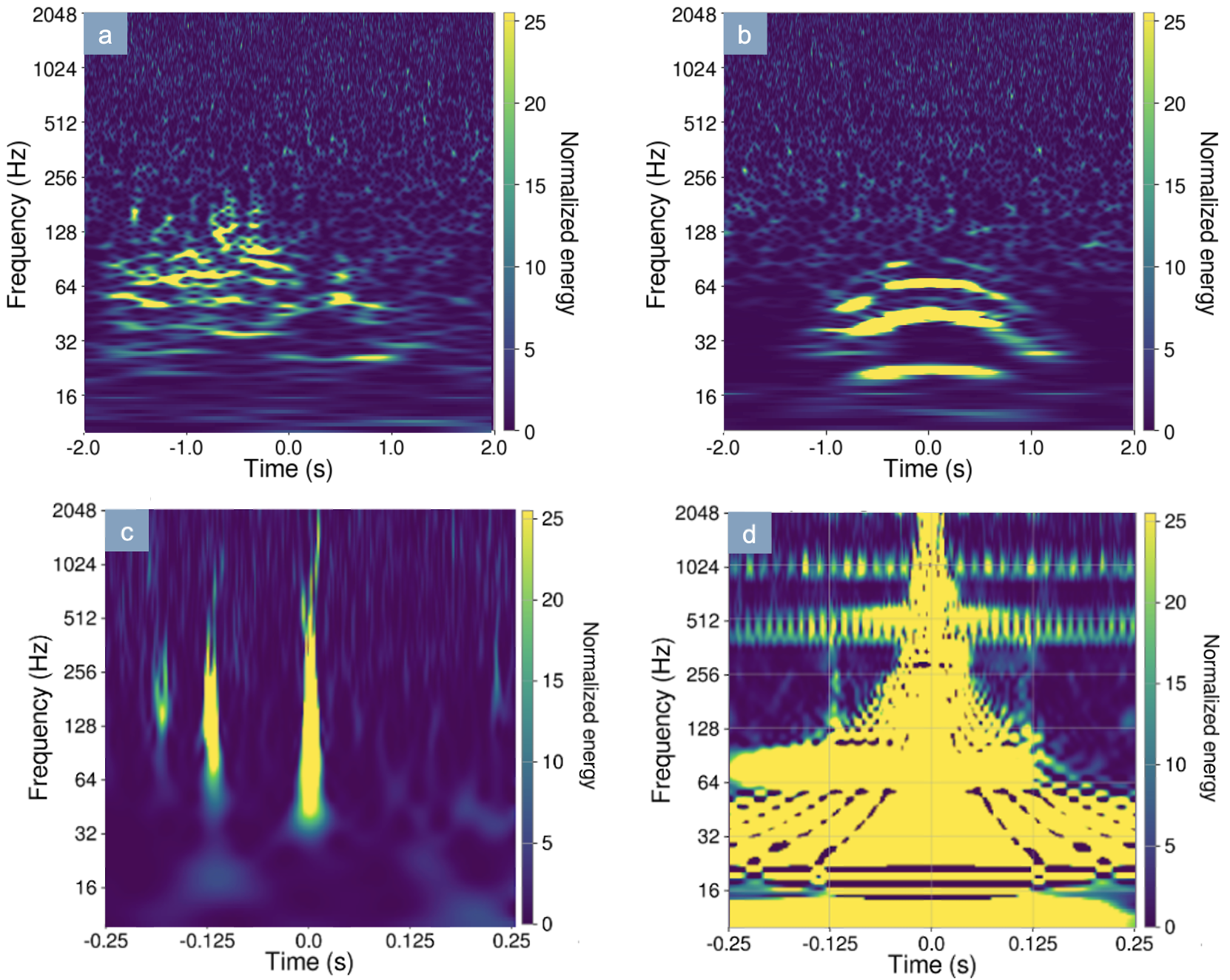}
    \caption{Examples of the glitches used for the validation study that tests glitches not included in the original GSpyNetTree training sets: (a) thunder glitch, (b) Scattering glitch, (c) Repeating Blips, and (d) Extremely Loud glitch. The Thunder and Scattering glitches are shown in a different time scale than the other glitches to highlight their morphology and duration.}
    \label{fig:exotic}
\end{figure}

Our second validation study focused on studying the CNNs' performance for glitches not included in the original training set, which was restricted to avoid excess glitch classes that could reduce CNN performance. We selected 8 samples of thunder glitches, all of which occurred during O3, and more than 100 samples of Scattering, Extremely Loud, and Repeating Blips glitches from previous observing runs. An example of each is shown in Figure \ref{fig:exotic}. While the former type of glitch is not a Gravity Spy class, the three latter are part of the Gravity Spy training set.

We selected each type of glitch to address possible classification challenges that may arise with GSpyNetTree during O4. Scattering and thunder glitches are fairly common \cite{derek}, so it is more likely that they may appear at the same time as a candidate GW. Repeating Blips were an interesting case because we already had both Blips and Low-frequency blips in the training sets of the three classifiers. Studying the CNNs' performance in cases where these glitches repeated in the spectrograms allowed us to preliminarily study how the CNN performed with multiple glitches in the same visualization. Extremely loud glitches are morphologically similar to Koi Fish glitches: they both extend in a wide frequency and time range. Thus, including them in our validation study allowed us to evaluate the CNNs performance on morphologically similar glitches. 

\begin{figure}[htbp]
\centering
\captionsetup[subfigure]{font=small}
     \begin{subfigure}[b]{0.52\textwidth}
         \includegraphics[width=\textwidth]{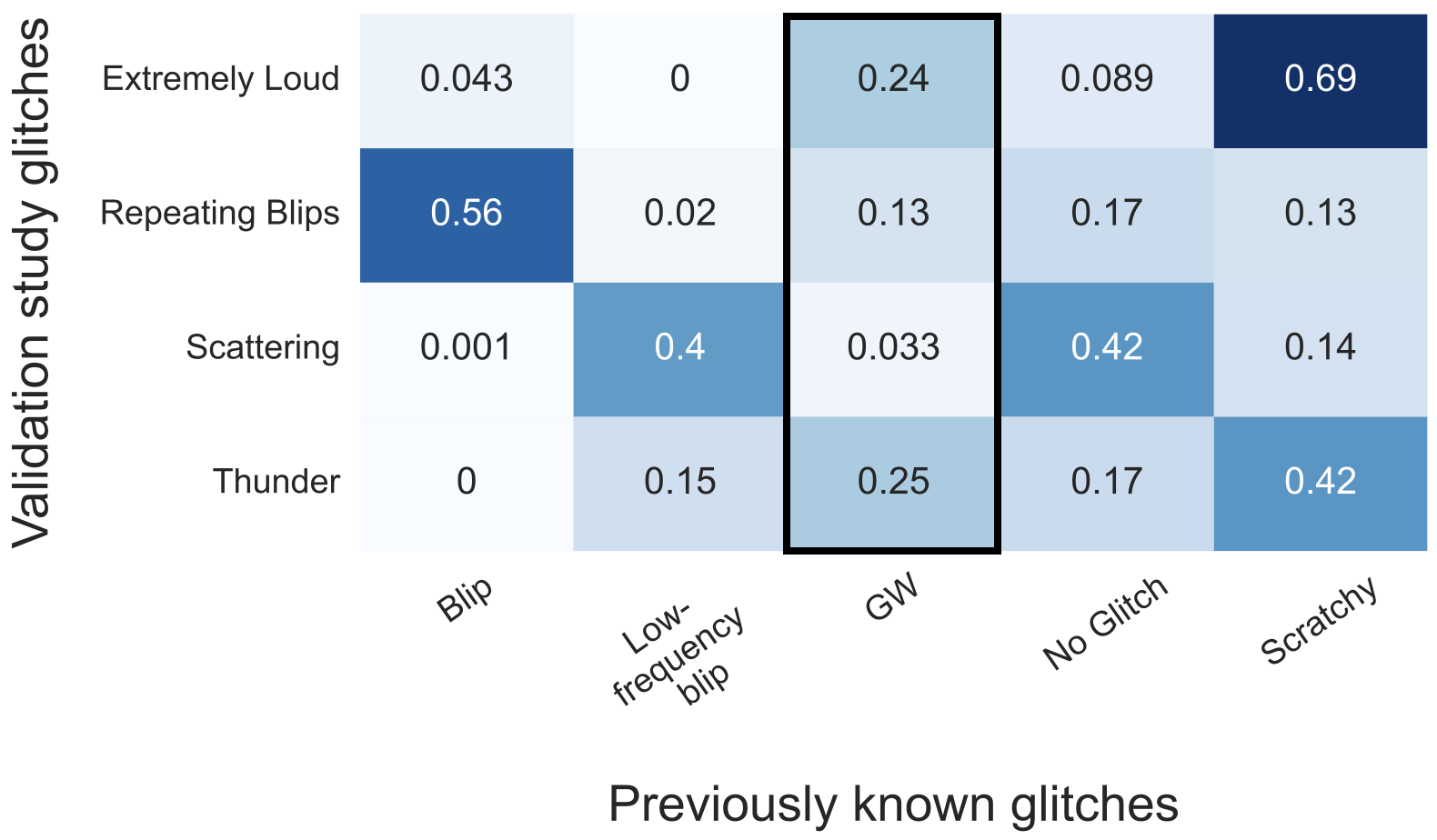}
         \caption[]{Confusion matrix for the LM classifier of GSpyNetTree for glitches not included in the original training set.} 
         \label{fig:exotic_lm}
     \end{subfigure}
     \hfill
     \begin{subfigure}[b]{0.42\textwidth}
         \includegraphics[width=\textwidth]{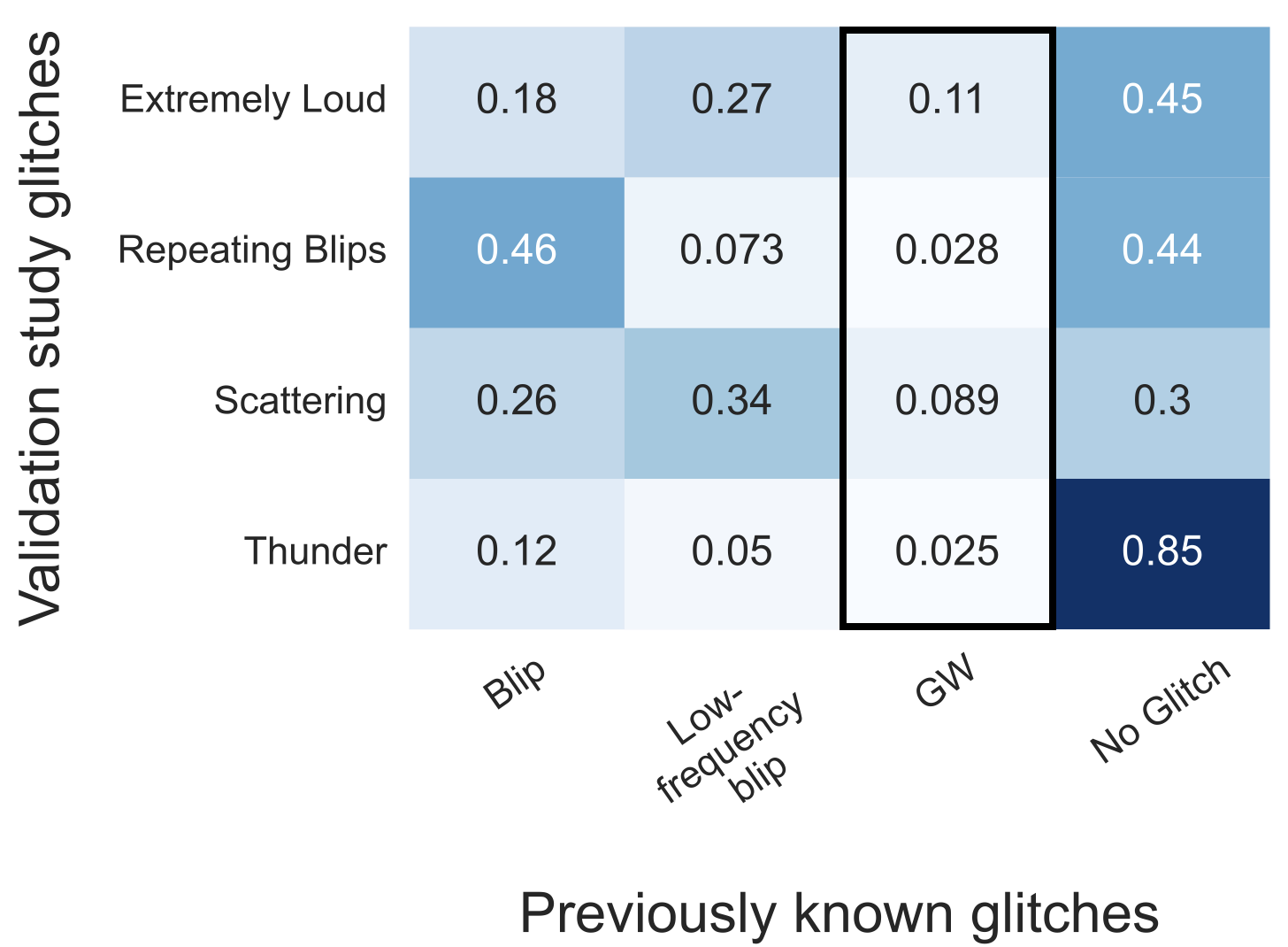}
         \caption[]{Confusion matrix for the EHM classifier of GSpyNetTree for glitches not included in the original training set.} 
         \label{fig:exotic_ehm}
     \end{subfigure}

     \vskip\baselineskip
     \begin{subfigure}[b]{0.6\textwidth}
         \includegraphics[width=\textwidth]{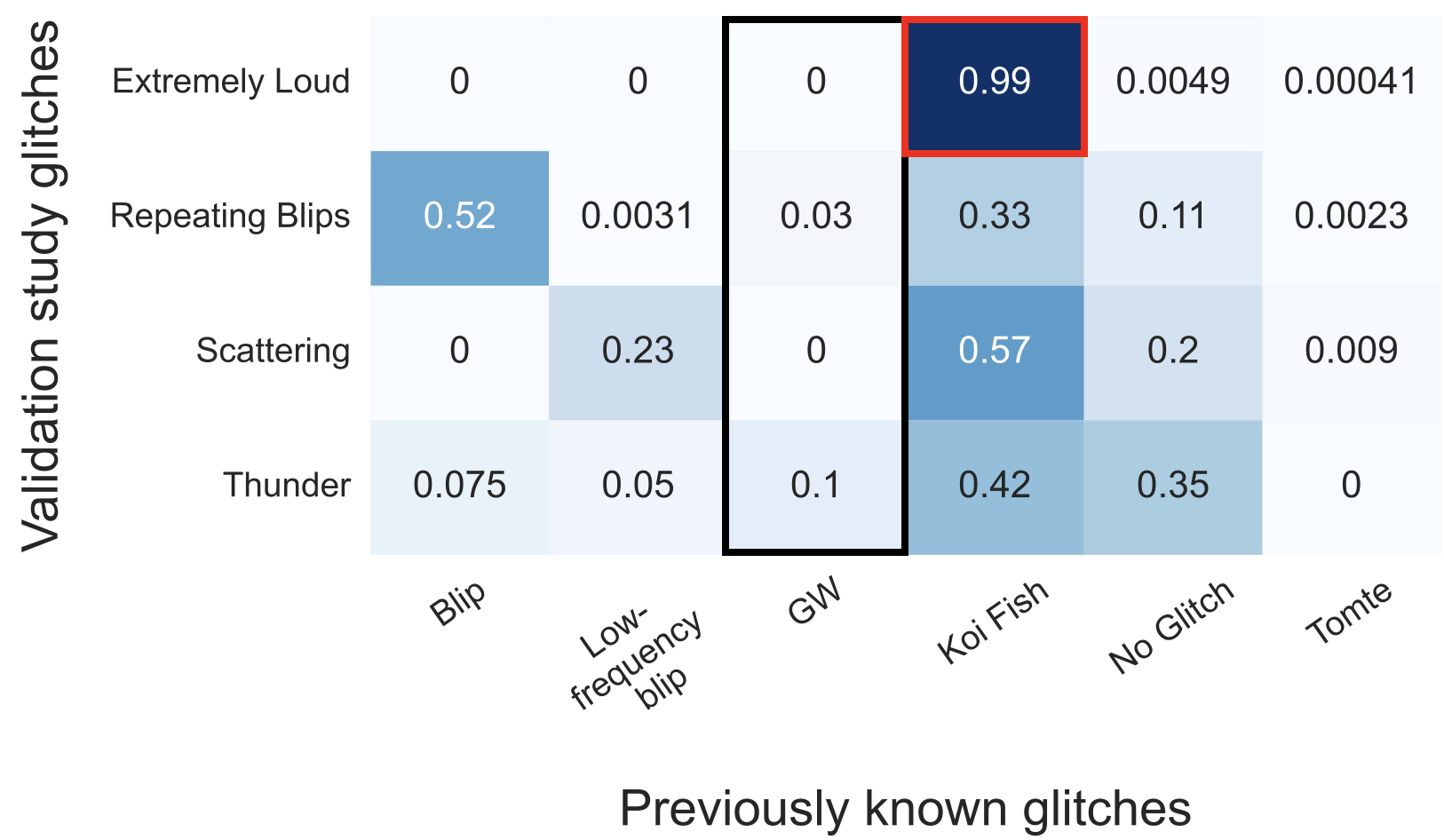}
         \caption[]{Confusion matrix for the HM classifier of GSpyNetTree for glitches not included in the original training set.} 
         \label{fig:exotic_hm}
     \end{subfigure}
     \hfill
        \caption{Confusion matrices for the (a) LM, (b) EHM, and (c) HM classifiers in the testing set for the validation study of glitches not included in the original training set: Scattering, Repeating Blips, Thunder, and Extremely Loud glitches. The x and y axes represent the predicted and true classes, respectively, and the confusion matrices are normalized by the total number of glitches of each class of the validation set. The black boxes highlight the fraction of glitches misclassified as GWs. The red box in the confusion matrix of the HM classifier highlights the fraction of Koi Fish glitches misclassified as Extremely Loud glitches.    } 
        \label{fig:exotic_cm}
\end{figure}

We tested the three GSpyNetTree classifiers using all the examples considered for this validation study. The classification results for the LM, HM, and EHM classifiers are shown in Figures \ref{fig:exotic_lm}, \ref{fig:exotic_hm}, and \ref{fig:exotic_ehm}, respectively. First, it is important to highlight that several glitches are often classified as GWs. A lack of robustness to glitches not included in the training set could be particularly problematic in O4, as new noise sources may appear (with the increase in the detectors' sensitivity). 

Overall, this validation study shows that the CNNs have low confidence when classifying new glitches. The classification probability is almost evenly distributed among all classes (including GWs) for most of the glitches in the three classifiers. This is, however, not the case for Repeating Blips and Scattering (in all three CNNs), and Extremely Loud glitches (in the HM classifier). Repeating Blips were classified as Blips with 56\%, 52\%, and 46\% accuracy for the LM, HM, and EHM classifiers, respectively. Even though these results can be further improved, this is promising evidence that the CNN architecture can be tuned to better classify repeating glitches. 

Scattering glitches have significantly different behavior in the three classifiers. In the LM CNN, they are mistaken for No Glitches (42\%) and Low-frequency blips (40\%), possibly because they evolve in a low-frequency range. However, the probability is distributed among the three non-GW classes in the EHM classifier. On the other hand, the HM CNN classifies most of the Scattering glitches (57\%) as Koi Fish glitches, suggesting that it focuses on the broad (low)-frequency range the Scattering glitches cover. Since Scattering glitches are one of the most common glitches in current GW detectors and having shown such different behavior among the three CNNs, this is strong evidence that this glitch class should be included as part of the O4-era GSpyNetTree training set. 

Moreover, morphologically similar glitches (particularly Koi Fish and Extremely Loud glitches, which both display saturated spectrograms) were classified with 99\% accuracy as the class already known by the HM CNN, as shown in the red box in Figure \ref{fig:exotic_hm}, and they are not easily mistaken for GWs. This is desirable as GSpyNetTree should minimize flagging non-astrophysical glitches as GW candidates. Even though not as morphologically similar as Koi Fish glitches, the LM CNN classified Extremely Loud glitches as the most dispersed in time and frequency glitch it knows: the Scratchy glitch. However, as opposed to the HM CNN, it misclassifies 24\% of them as GWs, which is undesirable if we intend to mitigate the false positive rate. On the other hand, the EHM classifier has a peculiar behavior when classifying both thunder and Extremely Loud glitches. As shown in Figure \ref{fig:exotic_ehm}, 85\% and 45\% of the samples are misclassified, respectively, as No Glitches. Covering a wide time-frequency range with a high SNR, it is counter-intuitive that they are both classified in the No Glitch class. 

In light of this behavior, an option to address misclassifications of new glitches is creating a new class for unknown or miscellaneous noise sources. However, this may reduce both the overall and GW classification accuracy, as we would be including glitches with substantially different morphologies in the same class, and it does not allow us to investigate new noise sources. Although the easiest solution is to include as many new classes in the CNNs as glitches arise, this approach will potentially reduce the performance of the CNN (as it has more classes it can get confused with). 

To solve this issue for the O4-era GSpyNetTree, we propose to use each CNN as a feature extractor in a semi/unsupervised learning task, similar to the approach followed by George et al. \cite{uiuc}. This way, each CNN's last layer would be projected to a 3-dimensional space using the 
T-distributed Stochastic Neighborhood Embedding (t-SNE) algorithm \cite{tsne}, such that glitches form clusters whose positions in the 3-d space depend on their morphology. Outliers (glitches not included in the original training set) may create a new cluster (i.e., a new glitch class) or may be separated enough to be segregated from training set classes for further detector characterization and glitch mitigation studies. Compared to the other alternatives, this approach would successfully distinguish new glitch classes from GWs, which is desirable for the O4-era GSpyNetTree signal-vs-glitch classifiers to maximize the detection of astrophysical events with a low false alarm rate. 

\subsection{Testing overlapping glitches and GWs }

As the aLIGO and AdVirgo detectors become more sensitive and the rate of detected events increases, the probability of overlapping glitches and GW signals in strain data also rises. This is a very likely scenario in O4, as it already happened with 26\% and 23\% of the candidates during the first \cite{o3a} and second \cite{o3b} parts of O3. Thus, we tested GSpyNetTree's performance in cases in which candidates occur in a similar time window as glitches to investigate whether this hinders the classification of true GWs. 

\begin{figure}[htbp]
  \centering
    \includegraphics[scale=0.35]{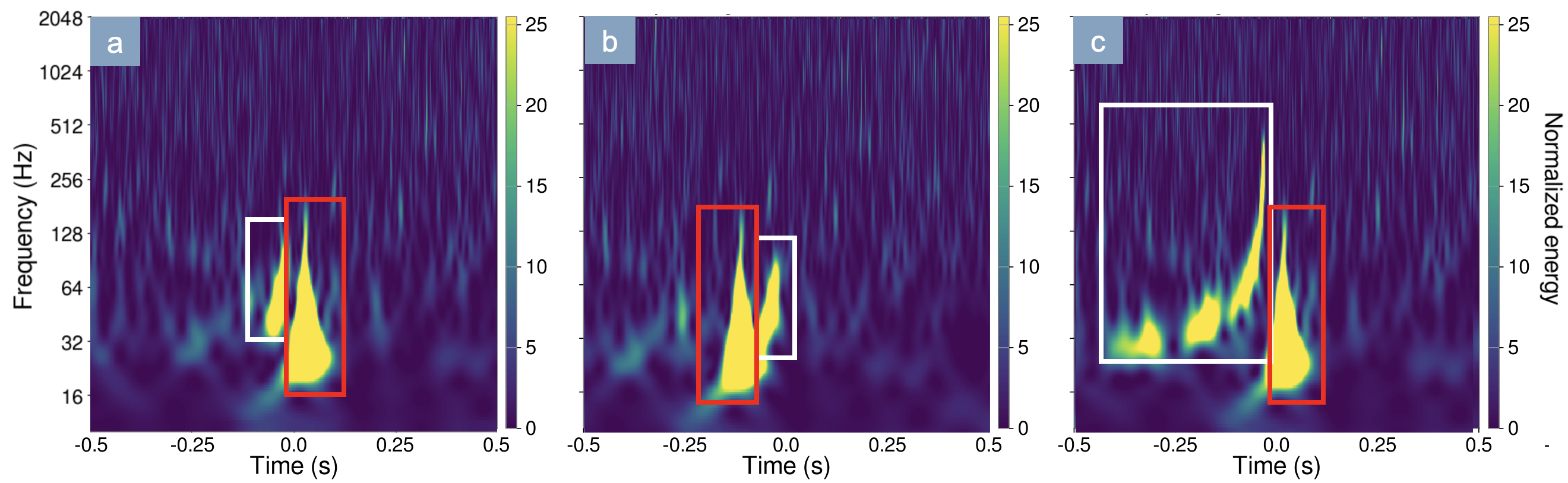}
    \caption{Samples of GWs (white boxes) injected in close proximity to Tomte glitches (red boxes). (a) and (b) show the same GW (with a total mass of 191.8 $\solarmass$), but the glitch has a different time-offset with respect to the GW: $0.011$ s and $-0.130$ s, respectively. (c) shows the Tomte glitch with the same time offset as in (a), but the GW signal has different astrophysical parameters (a total mass of 52 $\solarmass$). }
    \label{fig:tomte_high_mass}
\end{figure}

To test GSpyNetTree's ability to accurately classify GWs in the presence of glitches, we generated 30 events per glitch class, for each of the classifiers. First, we injected 3 GWs with different astrophysical parameters (but in the same parameter space as in training) into one of the glitches already learnt in training by the classifier. We then generated 10 normally distributed offsets ($\mu = 0\;, \sigma = 0.25\; \mathrm{s}$) so that the glitch was shifted in time with respect to the GW signal. This way, we could test GW signals (with different SNR, mass, and spin of the mergers) with offset glitches occurring in the same time window of each spectrogram. An example of a high mass GW and a Tomte glitch can be seen in Figure \ref{fig:tomte_high_mass}. 

\begin{figure}[htbp]
  \centering
    \includegraphics[scale=0.425]{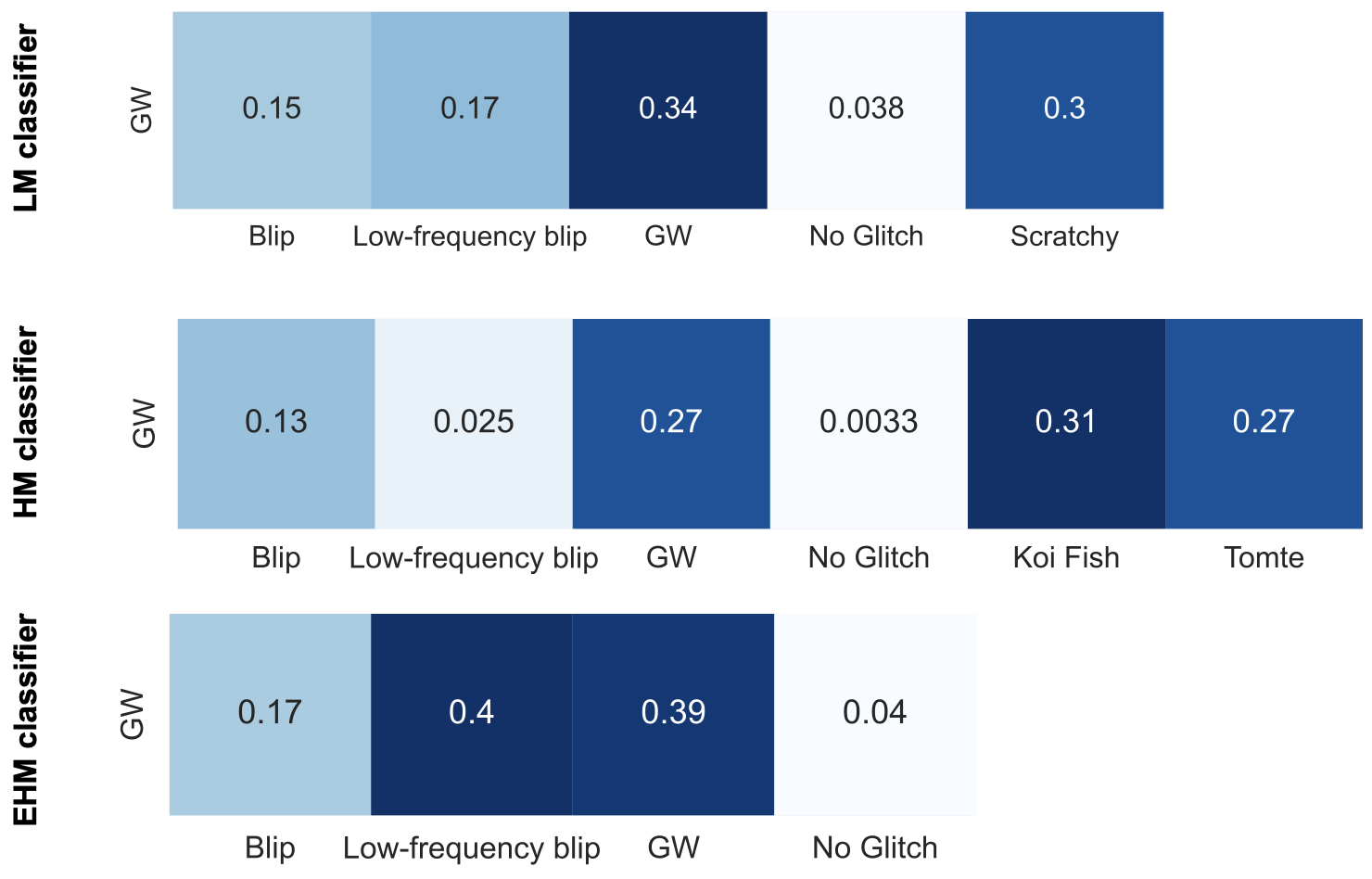}
    \caption{Accuracy results for the LM (top), HM (middle), and EHM (bottom) classifiers for the overlapping gravitational-waves and signals in the validation study. More than 60\% (70\% for the HM classifier) of the samples are misclassified as glitches.}
    \label{fig:results_g+s}
\end{figure}

Figure \ref{fig:results_g+s} shows the results of this validation study for the three classifiers. More than 60\% of the GW signals in the presence of glitches are misclassified as glitches (70\% for the HM classifier). Whenever not flagged as GWs, the LM and HM CNNs classified the overlapping GW and glitch events as the longest duration or most saturated glitches they were exposed to during training: Scratchy glitches and Blips in the LM classifier and Koi Fish and Tomte glitches in the HM classifier, as shown in Figure \ref{fig:results_g+s}.

The EHM classifier tends to inaccurately classify overlapping GWs and glitches as Low-frequency blips, with almost the same probability (40\%) as it classifies them as GWs (39\%). In all three CNNs, having such a high fraction of candidates flagged as glitches may result in unnecessarily vetoed candidates, an issue that must be avoided for O4. It is clear that GSpyNetTree's robustness to these events must be increased for it to be used as an element of fully-automated validation of LIGO-Virgo event candidates.

\begin{figure}[htbp]
  \centering
    \includegraphics[scale=0.425]{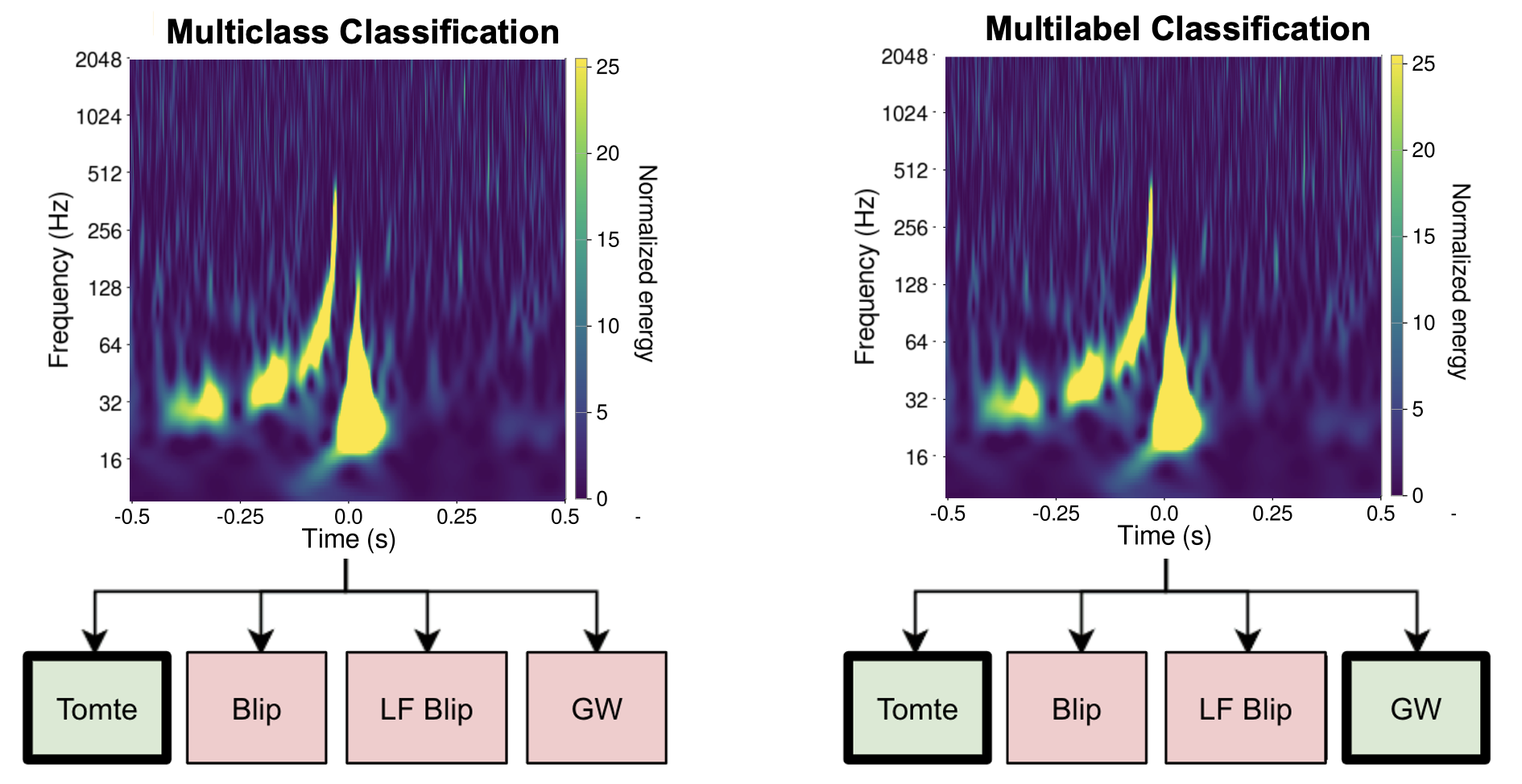}
    \caption{Differences between a multi-class (left) and a multi-label (right) classifier. In the first case, the CNN is only able to identify one of the mutually exclusive classes it was trained on (making it prone to misclassify astrophysical data as glitches). In this particular example, the classifier only detects the Tomte glitch from a sample with an overlapping GW signal. The multi-label architecture would solve this issue. Being able to output several classes from a single sample, both the GW signal and the Tomte glitch would be accurately classified.}
    \label{fig:multilabel}
\end{figure}

We expect GSpyNetTree misclassifies most of the overlapping glitch and GW events because it is a multi-class classifier: it outputs the probability that a given sample belongs to one particular class, which is independent and mutually exclusive with the other classes.  This justifies exploring a multi-label classifier, which would support the prediction of multiple mutually non-exclusive classes, also called labels \cite{geron}. This way, the GW signal (or, equivalently, the Tomte glitch) will not be ignored by the classifier, but both of them will be correctly classified as the classes they belong to, as shown in Figure \ref{fig:multilabel}. 

We expect that a multi-label classifier will also be able to accurately classify instances of overlapping glitches (such as the Repeating Blips, explained in section \ref{exotic} and shown in Figure \ref{fig:exotic}), without including them as a separate class during training. As GW detector data contains a high rate of glitches, glitches of different types and morphologies often overlap, and we expect that a future multi-label architecture of GSpyNetTree would also be able to accurately classify candidate events associated with overlapping glitches.

\section{Conclusions}

Based on the proof of principle introduced in Jarov et al. \cite{seraphim} and building on the success of Gravity Spy \cite{gspy}, we built GSpyNetTree, a multi-CNN decision tree of signal-vs-glitch classifiers. We leverage three CNNs, initially based on the original Gravity Spy architecture \cite{gspy}, each of which is trained with specialized training sets, depending on the mass of simulated GW mergers and morphologically similar glitches. We consider three total mass ranges: 3-50 \solarmass, 50-250 \solarmass, and 250-350 \solarmass\; for low, high, and extremely-high mass GWs, respectively. 

After achieving a 92\% accuracy for the GW class in the three CNNs, we noticed that we could use another architecture to further increase the amount of accurately classified samples. We implemented Inception V3, Google's state-of-the-art CNN for image classification tasks, and achieved more than 96\% accuracy for all classifiers. 

With better classification results, we tested GSpyNetTree for the O4-era detector's sensitivity, aiming to have all the signal-vs-glitch classifiers robust to a wide variety of background noise, new sources of glitches, and the likely occurrence of overlapping glitches and GWs. To test the CNNs' reliance on background, we evaluated GSpyNetTree's performance on the same types of glitches as in training, but this time using data with non-linear subtraction of 60 Hz AC power artifacts, as expected in O4. We discovered that the classification accuracy decreased to less than 77\% for the three classifiers, suggesting that even small changes in background noise significantly impact classification. For future work, we suggest creating a new training set that includes the original glitches and the glitches with the non-linear subtraction, as well as Virgo glitches and GWs, to mitigate dependence on background noise for the O4-era GSpyNetTree. 

We also tested GSpyNetTree's performance on glitches not included in the original training set: Scattering, Thunder, Repeating Blips, and Extremely Loud glitches. We noted that many non-astrophysical events were flagged as GW candidates, which is undesirable as we want to limit the false positive rate. Although the classification results varied considerably among the three CNNs, it was clear that very common glitches that are more likely to appear at the same time as a GW candidate (such as Scattering) should be included in GSpyNetTree's O4 training set. We observed that glitches morphologically similar to a training glitch class were confidently predicted as that class. Following George et al. \cite{uiuc}, we propose to add an unsupervised learning step that uses the CNNs as feature extractors projected into a 3-dimensional space, such that glitches form clusters where position depends on their morphology. New glitches with different morphologies might be segregated enough in that 3-d space from the original classes (especially true GWs), avoiding the need to increase the number of training classes used in the CNNs. 

The last validation study we conducted evaluated GSpyNetTree's current capabilities in the likely case of overlapping GW signals and glitches during O4. We noticed that, for the three CNNs, more than 60\% of the astrophysical events with glitches nearby in time were classified as glitches. This is undesirable for O4, as we want to avoid misclassifying GW candidates. To address this, we propose future efforts implement a multi-label architecture replacing the current classifiers. This would allow glitches and GW signals appearing in close proximity to be simultaneously and correctly classified.

In summary, we suggest that the O4-era GSpyNetTree consist of three Inception V3 multi-label classifiers, each trained on a specialized training set including GWs in the three total mass ranges previously studied, along with morphologically similar glitches. These training sets should incorporate data augmentation, apply the Mercator transform to EHM samples, and include LIGO glitches of previous observing runs, as well as examples with a non-linear subtraction of AC 60 Hz power artifacts and Virgo glitches, to account for different backgrounds. Lastly, we suggest using the CNNs as feature extractors to detect and cluster new glitch classes, as well as known glitch classes not included in the GSpyNetTree training set.

\section{Acknowledgements}

This material is based upon work supported by NSF's LIGO Laboratory which is a major facility fully funded by the National Science Foundation. We would like to acknowledge and thank the Detector Characterization team of the LSC for their comments and suggestions, particularly to Beverly Berger, Derek Davis, Seraphim Jarov, and Siddharth Soni for very useful discussions. We also thank ManLeong (Mervyn) Chan for his continuous technical support during this project. The authors are grateful for computational resources provided by the LIGO Laboratory and supported by National Science Foundation Grants PHY-0757058 and PHY-0823459. SA is supported by the Mitacs Globalink Research Internship, AL and JD by the NSERC USRA award, and JM by the Canada Research Chairs program. 
 
\newpage
\printbibliography 

@unpublished{seraphim,
  author = {{S. Jarov et al.}},
  title = {A new method to distinguish gravitational-wave
signals from detector glitches with {Gravity} {Spy}},
  year = {(in prep)},
  %eprint = {}
}

@INPROCEEDINGS{deeper,
  author={Szegedy, Christian and Wei Liu and Yangqing Jia and Sermanet, Pierre and Reed, Scott and Anguelov, Dragomir and Erhan, Dumitru and Vanhoucke, Vincent and Rabinovich, Andrew},
  booktitle={{2015 IEEE Conference on Computer Vision and Pattern Recognition (CVPR)}}, 
  title={Going deeper with convolutions}, 
  year={2015},
  volume={},
  number={},
  pages={1-9},
  doi={10.1109/CVPR.2015.7298594}}

@article{gspy,
doi = {10.1088/1361-6382/aa5cea},
url = {https://dx.doi.org/10.1088/1361-6382/aa5cea},
year = {2017},
month = {02},
publisher = {IOP Publishing},
volume = {34},
number = {6},
pages = {064003},
author = {M Zevin and S Coughlin and S Bahaadini and E Besler and N Rohani and S Allen and M Cabero and K Crowston and A K Katsaggelos and S L Larson and T K Lee and C Lintott and T B Littenberg and A Lundgren and C Østerlund and J R Smith and L Trouille and V Kalogera},
title = {{Gravity Spy: integrating advanced LIGO detector characterization, machine learning, and citizen science}},
journal = {Classical and Quantum Gravity}
}

@article{bahaadini,
	title = {Machine learning for {Gravity} {Spy}: {Glitch} classification and dataset},
	volume = {444},
	issn = {0020-0255},
	url = {https://www.sciencedirect.com/science/article/pii/S0020025518301634},
	doi = {https://doi.org/10.1016/j.ins.2018.02.068},
	journal = {Information Sciences},
	author = {Bahaadini, S. and Noroozi, V. and Rohani, N. and Coughlin, S. and Zevin, M. and Smith, J. R. and Kalogera, V. and Katsaggelos, A.},
	year = {2018},
	pages = {172--186},
}

@INPROCEEDINGS{dataset_size_performance,
  author={Luo, Chao and Li, Xiaojie and Wang, Lutao and He, Jia and Li, Denggao and Zhou, Jiliu},
  booktitle={{2018 5th International Conference on Systems and Informatics (ICSAI)}}, 
  title={{How Does the Data set Affect CNN-based Image Classification Performance?}}, 
  year={2018},
  volume={},
  number={},
  pages={361-366},
  doi={10.1109/ICSAI.2018.8599448}
 }

@misc{ligodv,
	title = {{LigoDV}-web: {Providing} easy, secure and universal access to a large distributed scientific data store for the {LIGO} {Scientific} {Collaboration}},
	shorttitle = {{LigoDV}-web},
	url = {http://arxiv.org/abs/1611.01089},
	doi = {10.48550/arXiv.1611.01089},
	urldate = {2022-11-18},
	publisher = {arXiv},
	author = {Areeda, Joseph S. and Smith, Joshua R. and Lundgren, Andrew P. and Maros, Edward and Macleod, Duncan M. and Zweizig, John},
	month = nov,
	year = {2016},
	note = {arXiv:1611.01089 [astro-ph, physics:gr-qc]},
}

@misc{lalsuite,
       author         = "{The LIGO Scientific Collaboration}",
       title          = "{LIGO} {A}lgorithm {L}ibrary - {LALS}uite",
       howpublished   = "free software (GPL)",
       doi            = "10.7935/GT1W-FZ16",
       year           = "2018"
 }

@misc{cnn,
	title = {An {Introduction} to {Convolutional} {Neural} {Networks}},
	url = {http://arxiv.org/abs/1511.08458},
	doi = {10.48550/arXiv.1511.08458},
	urldate = {2022-11-18},
	publisher = {arXiv},
	author = {O'Shea, Keiron and Nash, Ryan},
	month = dec,
	year = {2015},
	note = {arXiv:1511.08458 [cs]},
}

@INPROCEEDINGS{inceptionv3,
  author={Szegedy, Christian and Vanhoucke, Vincent and Ioffe, Sergey and Shlens, Jon and Wojna, Zbigniew},
  booktitle={{2016 IEEE Conference on Computer Vision and Pattern Recognition (CVPR)}}, 
  title={{Rethinking the Inception Architecture for Computer Vision}}, 
  year={2016},
  volume={},
  number={},
  pages={2818-2826},
  doi={10.1109/CVPR.2016.308}}

@inproceedings{uiuc,
author = {George, Daniel and Shen, Hongyu and Huerta, Eliu},
title = {{Glitch Classification and Clustering for LIGO with Deep Transfer Learning}},
booktitle = "{NiPS Summer School 2017}",
eprint = "1711.07468",
archivePrefix = "arXiv",
primaryClass = "astro-ph.IM",
month = "11",
year = "2017"
}

@article{tsne,
  title={{Visualizing data using t-SNE.}},
  author={Van der Maaten, Laurens and Hinton, Geoffrey},
  journal={Journal of machine learning research},
  volume={9},
  number={11},
  year={2008}
}

@article{jane,
	doi = {10.1088/1361-6382/acb633},
	url = {https://doi.org/10.1088\%2F1361-6382\%2Facb633},
	year = 2023,
	month = {02},
	publisher = {{IOP} Publishing},
	volume = {40},
	number = {6},
	pages = {065004},
	author = {J Glanzer and S Banagiri and S B Coughlin and S Soni and M Zevin and C P L Berry and O Patane and S Bahaadini and N Rohani and K Crowston and V Kalogera and C {\O}sterlund and L Trouille and A Katsaggelos},
	title = {{Data quality up to the third observing run of advanced {LIGO}: Gravity Spy glitch classifications}},
	journal = {Classical and Quantum Gravity}}

@article{1stdiscovery,
  title = {{Observation of Gravitational Waves from a Binary Black Hole Merger}},
  author = {Abbott, B. P. and Abbott, R. and Abbott, T. D. and Abernathy, M. R. and Acernese, F. and Ackley, K. and Adams, C. and Adams, T. and Addesso, P. and Adhikari, R. X. and Adya, V. B. and Affeldt, C. and Agathos, M. and Agatsuma, K. and Aggarwal, N. and Aguiar, O. D. and Aiello, L. and Ain, A. and Ajith, P. and Allen, B. and Allocca, A. and Altin, P. A. and Anderson, S. B. and Anderson, W. G. and Arai, K. and Arain, M. A. and Araya, M. C. and Arceneaux, C. C. and Areeda, J. S. and Arnaud, N. and Arun, K. G. and Ascenzi, S. and Ashton, G. and Ast, M. and Aston, S. M. and Astone, P. and Aufmuth, P. and Aulbert, C. and Babak, S. and Bacon, P. and Bader, M. K. M. and Baker, P. T. and Baldaccini, F. and Ballardin, G. and Ballmer, S. W. and Barayoga, J. C. and Barclay, S. E. and Barish, B. C. and Barker, D. and Barone, F. and Barr, B. and Barsotti, L. and Barsuglia, M. and Barta, D. and Bartlett, J. and Barton, M. A. and Bartos, I. and Bassiri, R. and Basti, A. and Batch, J. C. and Baune, C. and Bavigadda, V. and Bazzan, M. and Behnke, B. and Bejger, M. and Belczynski, C. and Bell, A. S. and Bell, C. J. and Berger, B. K. and Bergman, J. and Bergmann, G. and Berry, C. P. L. and Bersanetti, D. and Bertolini, A. and Betzwieser, J. and Bhagwat, S. and Bhandare, R. and Bilenko, I. A. and Billingsley, G. and Birch, J. and Birney, R. and Birnholtz, O. and Biscans, S. and Bisht, A. and Bitossi, M. and Biwer, C. and Bizouard, M. A. and Blackburn, J. K. and Blair, C. D. and Blair, D. G. and Blair, R. M. and Bloemen, S. and Bock, O. and Bodiya, T. P. and Boer, M. and Bogaert, G. and Bogan, C. and Bohe, A. and Bojtos, P. and Bond, C. and Bondu, F. and Bonnand, R. and Boom, B. A. and Bork, R. and Boschi, V. and Bose, S. and Bouffanais, Y. and Bozzi, A. and Bradaschia, C. and Brady, P. R. and Braginsky, V. B. and Branchesi, M. and Brau, J. E. and Briant, T. and Brillet, A. and Brinkmann, M. and Brisson, V. and Brockill, P. and Brooks, A. F. and Brown, D. A. and Brown, D. D. and Brown, N. M. and Buchanan, C. C. and Buikema, A. and Bulik, T. and Bulten, H. J. and Buonanno, A. and Buskulic, D. and Buy, C. and Byer, R. L. and Cabero, M. and Cadonati, L. and Cagnoli, G. and Cahillane, C. and Bustillo, J. Calder\'on and Callister, T. and Calloni, E. and Camp, J. B. and Cannon, K. C. and Cao, J. and Capano, C. D. and Capocasa, E. and Carbognani, F. and Caride, S. and Diaz, J. Casanueva and Casentini, C. and Caudill, S. and Cavagli\`a, M. and Cavalier, F. and Cavalieri, R. and Cella, G. and Cepeda, C. B. and Baiardi, L. Cerboni and Cerretani, G. and Cesarini, E. and Chakraborty, R. and Chalermsongsak, T. and Chamberlin, S. J. and Chan, M. and Chao, S. and Charlton, P. and Chassande-Mottin, E. and Chen, H. Y. and Chen, Y. and Cheng, C. and Chincarini, A. and Chiummo, A. and Cho, H. S. and Cho, M. and Chow, J. H. and Christensen, N. and Chu, Q. and Chua, S. and Chung, S. and Ciani, G. and Clara, F. and Clark, J. A. and Cleva, F. and Coccia, E. and Cohadon, P.-F. and Colla, A. and Collette, C. G. and Cominsky, L. and Constancio, M. and Conte, A. and Conti, L. and Cook, D. and Corbitt, T. R. and Cornish, N. and Corsi, A. and Cortese, S. and Costa, C. A. and Coughlin, M. W. and Coughlin, S. B. and Coulon, J.-P. and Countryman, S. T. and Couvares, P. and Cowan, E. E. and Coward, D. M. and Cowart, M. J. and Coyne, D. C. and Coyne, R. and Craig, K. and Creighton, J. D. E. and Creighton, T. D. and Cripe, J. and Crowder, S. G. and Cruise, A. M. and Cumming, A. and Cunningham, L. and Cuoco, E. and Canton, T. Dal and Danilishin, S. L. and D'Antonio, S. and Danzmann, K. and Darman, N. S. and Da Silva Costa, C. F. and Dattilo, V. and Dave, I. and Daveloza, H. P. and Davier, M. and Davies, G. S. and Daw, E. J. and Day, R. and De, S. and DeBra, D. and Debreczeni, G. and Degallaix, J. and De Laurentis, M. and Del\'eglise, S. and Del Pozzo, W. and Denker, T. and Dent, T. and Dereli, H. and Dergachev, V. and DeRosa, R. T. and De Rosa, R. and DeSalvo, R. and Dhurandhar, S. and D\'{\i}az, M. C. and Di Fiore, L. and Di Giovanni, M. and Di Lieto, A. and Di Pace, S. and Di Palma, I. and Di Virgilio, A. and Dojcinoski, G. and Dolique, V. and Donovan, F. and Dooley, K. L. and Doravari, S. and Douglas, R. and Downes, T. P. and Drago, M. and Drever, R. W. P. and Driggers, J. C. and Du, Z. and Ducrot, M. and Dwyer, S. E. and Edo, T. B. and Edwards, M. C. and Effler, A. and Eggenstein, H.-B. and Ehrens, P. and Eichholz, J. and Eikenberry, S. S. and Engels, W. and Essick, R. C. and Etzel, T. and Evans, M. and Evans, T. M. and Everett, R. and Factourovich, M. and Fafone, V. and Fair, H. and Fairhurst, S. and Fan, X. and Fang, Q. and Farinon, S. and Farr, B. and Farr, W. M. and Favata, M. and Fays, M. and Fehrmann, H. and Fejer, M. M. and Feldbaum, D. and Ferrante, I. and Ferreira, E. C. and Ferrini, F. and Fidecaro, F. and Finn, L. S. and Fiori, I. and Fiorucci, D. and Fisher, R. P. and Flaminio, R. and Fletcher, M. and Fong, H. and Fournier, J.-D. and Franco, S. and Frasca, S. and Frasconi, F. and Frede, M. and Frei, Z. and Freise, A. and Frey, R. and Frey, V. and Fricke, T. T. and Fritschel, P. and Frolov, V. V. and Fulda, P. and Fyffe, M. and Gabbard, H. A. G. and Gair, J. R. and Gammaitoni, L. and Gaonkar, S. G. and Garufi, F. and Gatto, A. and Gaur, G. and Gehrels, N. and Gemme, G. and Gendre, B. and Genin, E. and Gennai, A. and George, J. and Gergely, L. and Germain, V. and Ghosh, Abhirup and Ghosh, Archisman and Ghosh, S. and Giaime, J. A. and Giardina, K. D. and Giazotto, A. and Gill, K. and Glaefke, A. and Gleason, J. R. and Goetz, E. and Goetz, R. and Gondan, L. and Gonz\'alez, G. and Castro, J. M. Gonzalez and Gopakumar, A. and Gordon, N. A. and Gorodetsky, M. L. and Gossan, S. E. and Gosselin, M. and Gouaty, R. and Graef, C. and Graff, P. B. and Granata, M. and Grant, A. and Gras, S. and Gray, C. and Greco, G. and Green, A. C. and Greenhalgh, R. J. S. and Groot, P. and Grote, H. and Grunewald, S. and Guidi, G. M. and Guo, X. and Gupta, A. and Gupta, M. K. and Gushwa, K. E. and Gustafson, E. K. and Gustafson, R. and Hacker, J. J. and Hall, B. R. and Hall, E. D. and Hammond, G. and Haney, M. and Hanke, M. M. and Hanks, J. and Hanna, C. and Hannam, M. D. and Hanson, J. and Hardwick, T. and Harms, J. and Harry, G. M. and Harry, I. W. and Hart, M. J. and Hartman, M. T. and Haster, C.-J. and Haughian, K. and Healy, J. and Heefner, J. and Heidmann, A. and Heintze, M. C. and Heinzel, G. and Heitmann, H. and Hello, P. and Hemming, G. and Hendry, M. and Heng, I. S. and Hennig, J. and Heptonstall, A. W. and Heurs, M. and Hild, S. and Hoak, D. and Hodge, K. A. and Hofman, D. and Hollitt, S. E. and Holt, K. and Holz, D. E. and Hopkins, P. and Hosken, D. J. and Hough, J. and Houston, E. A. and Howell, E. J. and Hu, Y. M. and Huang, S. and Huerta, E. A. and Huet, D. and Hughey, B. and Husa, S. and Huttner, S. H. and Huynh-Dinh, T. and Idrisy, A. and Indik, N. and Ingram, D. R. and Inta, R. and Isa, H. N. and Isac, J.-M. and Isi, M. and Islas, G. and Isogai, T. and Iyer, B. R. and Izumi, K. and Jacobson, M. B. and Jacqmin, T. and Jang, H. and Jani, K. and Jaranowski, P. and Jawahar, S. and Jim\'enez-Forteza, F. and Johnson, W. W. and Johnson-McDaniel, N. K. and Jones, D. I. and Jones, R. and Jonker, R. J. G. and Ju, L. and Haris, K. and Kalaghatgi, C. V. and Kalogera, V. and Kandhasamy, S. and Kang, G. and Kanner, J. B. and Karki, S. and Kasprzack, M. and Katsavounidis, E. and Katzman, W. and Kaufer, S. and Kaur, T. and Kawabe, K. and Kawazoe, F. and K\'ef\'elian, F. and Kehl, M. S. and Keitel, D. and Kelley, D. B. and Kells, W. and Kennedy, R. and Keppel, D. G. and Key, J. S. and Khalaidovski, A. and Khalili, F. Y. and Khan, I. and Khan, S. and Khan, Z. and Khazanov, E. A. and Kijbunchoo, N. and Kim, C. and Kim, J. and Kim, K. and Kim, Nam-Gyu and Kim, Namjun and Kim, Y.-M. and King, E. J. and King, P. J. and Kinzel, D. L. and Kissel, J. S. and Kleybolte, L. and Klimenko, S. and Koehlenbeck, S. M. and Kokeyama, K. and Koley, S. and Kondrashov, V. and Kontos, A. and Koranda, S. and Korobko, M. and Korth, W. Z. and Kowalska, I. and Kozak, D. B. and Kringel, V. and Krishnan, B. and Kr\'olak, A. and Krueger, C. and Kuehn, G. and Kumar, P. and Kumar, R. and Kuo, L. and Kutynia, A. and Kwee, P. and Lackey, B. D. and Landry, M. and Lange, J. and Lantz, B. and Lasky, P. D. and Lazzarini, A. and Lazzaro, C. and Leaci, P. and Leavey, S. and Lebigot, E. O. and Lee, C. H. and Lee, H. K. and Lee, H. M. and Lee, K. and Lenon, A. and Leonardi, M. and Leong, J. R. and Leroy, N. and Letendre, N. and Levin, Y. and Levine, B. M. and Li, T. G. F. and Libson, A. and Littenberg, T. B. and Lockerbie, N. A. and Logue, J. and Lombardi, A. L. and London, L. T. and Lord, J. E. and Lorenzini, M. and Loriette, V. and Lormand, M. and Losurdo, G. and Lough, J. D. and Lousto, C. O. and Lovelace, G. and L\"uck, H. and Lundgren, A. P. and Luo, J. and Lynch, R. and Ma, Y. and MacDonald, T. and Machenschalk, B. and MacInnis, M. and Macleod, D. M. and Maga\~na-Sandoval, F. and Magee, R. M. and Mageswaran, M. and Majorana, E. and Maksimovic, I. and Malvezzi, V. and Man, N. and Mandel, I. and Mandic, V. and Mangano, V. and Mansell, G. L. and Manske, M. and Mantovani, M. and Marchesoni, F. and Marion, F. and M\'arka, S. and M\'arka, Z. and Markosyan, A. S. and Maros, E. and Martelli, F. and Martellini, L. and Martin, I. W. and Martin, R. M. and Martynov, D. V. and Marx, J. N. and Mason, K. and Masserot, A. and Massinger, T. J. and Masso-Reid, M. and Matichard, F. and Matone, L. and Mavalvala, N. and Mazumder, N. and Mazzolo, G. and McCarthy, R. and McClelland, D. E. and McCormick, S. and McGuire, S. C. and McIntyre, G. and McIver, J. and McManus, D. J. and McWilliams, S. T. and Meacher, D. and Meadors, G. D. and Meidam, J. and Melatos, A. and Mendell, G. and Mendoza-Gandara, D. and Mercer, R. A. and Merilh, E. and Merzougui, M. and Meshkov, S. and Messenger, C. and Messick, C. and Meyers, P. M. and Mezzani, F. and Miao, H. and Michel, C. and Middleton, H. and Mikhailov, E. E. and Milano, L. and Miller, J. and Millhouse, M. and Minenkov, Y. and Ming, J. and Mirshekari, S. and Mishra, C. and Mitra, S. and Mitrofanov, V. P. and Mitselmakher, G. and Mittleman, R. and Moggi, A. and Mohan, M. and Mohapatra, S. R. P. and Montani, M. and Moore, B. C. and Moore, C. J. and Moraru, D. and Moreno, G. and Morriss, S. R. and Mossavi, K. and Mours, B. and Mow-Lowry, C. M. and Mueller, C. L. and Mueller, G. and Muir, A. W. and Mukherjee, Arunava and Mukherjee, D. and Mukherjee, S. and Mukund, N. and Mullavey, A. and Munch, J. and Murphy, D. J. and Murray, P. G. and Mytidis, A. and Nardecchia, I. and Naticchioni, L. and Nayak, R. K. and Necula, V. and Nedkova, K. and Nelemans, G. and Neri, M. and Neunzert, A. and Newton, G. and Nguyen, T. T. and Nielsen, A. B. and Nissanke, S. and Nitz, A. and Nocera, F. and Nolting, D. and Normandin, M. E. N. and Nuttall, L. K. and Oberling, J. and Ochsner, E. and O'Dell, J. and Oelker, E. and Ogin, G. H. and Oh, J. J. and Oh, S. H. and Ohme, F. and Oliver, M. and Oppermann, P. and Oram, Richard J. and O'Reilly, B. and O'Shaughnessy, R. and Ott, C. D. and Ottaway, D. J. and Ottens, R. S. and Overmier, H. and Owen, B. J. and Pai, A. and Pai, S. A. and Palamos, J. R. and Palashov, O. and Palomba, C. and Pal-Singh, A. and Pan, H. and Pan, Y. and Pankow, C. and Pannarale, F. and Pant, B. C. and Paoletti, F. and Paoli, A. and Papa, M. A. and Paris, H. R. and Parker, W. and Pascucci, D. and Pasqualetti, A. and Passaquieti, R. and Passuello, D. and Patricelli, B. and Patrick, Z. and Pearlstone, B. L. and Pedraza, M. and Pedurand, R. and Pekowsky, L. and Pele, A. and Penn, S. and Perreca, A. and Pfeiffer, H. P. and Phelps, M. and Piccinni, O. and Pichot, M. and Pickenpack, M. and Piergiovanni, F. and Pierro, V. and Pillant, G. and Pinard, L. and Pinto, I. M. and Pitkin, M. and Poeld, J. H. and Poggiani, R. and Popolizio, P. and Post, A. and Powell, J. and Prasad, J. and Predoi, V. and Premachandra, S. S. and Prestegard, T. and Price, L. R. and Prijatelj, M. and Principe, M. and Privitera, S. and Prix, R. and Prodi, G. A. and Prokhorov, L. and Puncken, O. and Punturo, M. and Puppo, P. and P\"urrer, M. and Qi, H. and Qin, J. and Quetschke, V. and Quintero, E. A. and Quitzow-James, R. and Raab, F. J. and Rabeling, D. S. and Radkins, H. and Raffai, P. and Raja, S. and Rakhmanov, M. and Ramet, C. R. and Rapagnani, P. and Raymond, V. and Razzano, M. and Re, V. and Read, J. and Reed, C. M. and Regimbau, T. and Rei, L. and Reid, S. and Reitze, D. H. and Rew, H. and Reyes, S. D. and Ricci, F. and Riles, K. and Robertson, N. A. and Robie, R. and Robinet, F. and Rocchi, A. and Rolland, L. and Rollins, J. G. and Roma, V. J. and Romano, J. D. and Romano, R. and Romanov, G. and Romie, J. H. and Rosi\ifmmode \acute{n}\else \'{n}\fi{}ska, D. and Rowan, S. and R\"udiger, A. and Ruggi, P. and Ryan, K. and Sachdev, S. and Sadecki, T. and Sadeghian, L. and Salconi, L. and Saleem, M. and Salemi, F. and Samajdar, A. and Sammut, L. and Sampson, L. M. and Sanchez, E. J. and Sandberg, V. and Sandeen, B. and Sanders, G. H. and Sanders, J. R. and Sassolas, B. and Sathyaprakash, B. S. and Saulson, P. R. and Sauter, O. and Savage, R. L. and Sawadsky, A. and Schale, P. and Schilling, R. and Schmidt, J. and Schmidt, P. and Schnabel, R. and Schofield, R. M. S. and Sch\"onbeck, A. and Schreiber, E. and Schuette, D. and Schutz, B. F. and Scott, J. and Scott, S. M. and Sellers, D. and Sengupta, A. S. and Sentenac, D. and Sequino, V. and Sergeev, A. and Serna, G. and Setyawati, Y. and Sevigny, A. and Shaddock, D. A. and Shaffer, T. and Shah, S. and Shahriar, M. S. and Shaltev, M. and Shao, Z. and Shapiro, B. and Shawhan, P. and Sheperd, A. and Shoemaker, D. H. and Shoemaker, D. M. and Siellez, K. and Siemens, X. and Sigg, D. and Silva, A. D. and Simakov, D. and Singer, A. and Singer, L. P. and Singh, A. and Singh, R. and Singhal, A. and Sintes, A. M. and Slagmolen, B. J. J. and Smith, J. R. and Smith, M. R. and Smith, N. D. and Smith, R. J. E. and Son, E. J. and Sorazu, B. and Sorrentino, F. and Souradeep, T. and Srivastava, A. K. and Staley, A. and Steinke, M. and Steinlechner, J. and Steinlechner, S. and Steinmeyer, D. and Stephens, B. C. and Stevenson, S. P. and Stone, R. and Strain, K. A. and Straniero, N. and Stratta, G. and Strauss, N. A. and Strigin, S. and Sturani, R. and Stuver, A. L. and Summerscales, T. Z. and Sun, L. and Sutton, P. J. and Swinkels, B. L. and Szczepa\ifmmode \acute{n}\else \'{n}\fi{}czyk, M. J. and Tacca, M. and Talukder, D. and Tanner, D. B. and T\'apai, M. and Tarabrin, S. P. and Taracchini, A. and Taylor, R. and Theeg, T. and Thirugnanasambandam, M. P. and Thomas, E. G. and Thomas, M. and Thomas, P. and Thorne, K. A. and Thorne, K. S. and Thrane, E. and Tiwari, S. and Tiwari, V. and Tokmakov, K. V. and Tomlinson, C. and Tonelli, M. and Torres, C. V. and Torrie, C. I. and T\"oyr\"a, D. and Travasso, F. and Traylor, G. and Trifir\`o, D. and Tringali, M. C. and Trozzo, L. and Tse, M. and Turconi, M. and Tuyenbayev, D. and Ugolini, D. and Unnikrishnan, C. S. and Urban, A. L. and Usman, S. A. and Vahlbruch, H. and Vajente, G. and Valdes, G. and Vallisneri, M. and van Bakel, N. and van Beuzekom, M. and van den Brand, J. F. J. and Van Den Broeck, C. and Vander-Hyde, D. C. and van der Schaaf, L. and van Heijningen, J. V. and van Veggel, A. A. and Vardaro, M. and Vass, S. and Vas\'uth, M. and Vaulin, R. and Vecchio, A. and Vedovato, G. and Veitch, J. and Veitch, P. J. and Venkateswara, K. and Verkindt, D. and Vetrano, F. and Vicer\'e, A. and Vinciguerra, S. and Vine, D. J. and Vinet, J.-Y. and Vitale, S. and Vo, T. and Vocca, H. and Vorvick, C. and Voss, D. and Vousden, W. D. and Vyatchanin, S. P. and Wade, A. R. and Wade, L. E. and Wade, M. and Waldman, S. J. and Walker, M. and Wallace, L. and Walsh, S. and Wang, G. and Wang, H. and Wang, M. and Wang, X. and Wang, Y. and Ward, H. and Ward, R. L. and Warner, J. and Was, M. and Weaver, B. and Wei, L.-W. and Weinert, M. and Weinstein, A. J. and Weiss, R. and Welborn, T. and Wen, L. and We\ss{}els, P. and Westphal, T. and Wette, K. and Whelan, J. T. and Whitcomb, S. E. and White, D. J. and Whiting, B. F. and Wiesner, K. and Wilkinson, C. and Willems, P. A. and Williams, L. and Williams, R. D. and Williamson, A. R. and Willis, J. L. and Willke, B. and Wimmer, M. H. and Winkelmann, L. and Winkler, W. and Wipf, C. C. and Wiseman, A. G. and Wittel, H. and Woan, G. and Worden, J. and Wright, J. L. and Wu, G. and Yablon, J. and Yakushin, I. and Yam, W. and Yamamoto, H. and Yancey, C. C. and Yap, M. J. and Yu, H. and Yvert, M. and Zadro\ifmmode \dot{z}\else \.{z}\fi{}ny, A. and Zangrando, L. and Zanolin, M. and Zendri, J.-P. and Zevin, M. and Zhang, F. and Zhang, L. and Zhang, M. and Zhang, Y. and Zhao, C. and Zhou, M. and Zhou, Z. and Zhu, X. J. and Zucker, M. E. and Zuraw, S. E. and Zweizig, J.},
  collaboration = {LIGO Scientific Collaboration and Virgo Collaboration},
  journal = {Phys. Rev. Lett.},
  volume = {116},
  issue = {6},
  pages = {061102},
  numpages = {16},
  year = {2016},
  month = {02},
  publisher = {American Physical Society},
  doi = {10.1103/PhysRevLett.116.061102},
  url = {https://link.aps.org/doi/10.1103/PhysRevLett.116.061102}
}

@article{o1calib,
  title = {{Calibration of the Advanced LIGO detectors for the discovery of the binary black-hole merger GW150914}},
  author = {Abbott, B. P. and Abbott, R. and Abbott, T. D. and Abernathy, M. R. and Ackley, K. and Adams, C. and Addesso, P. and Adhikari, R. X. and Adya, V. B. and Affeldt, C. and Aggarwal, N. and Aguiar, O. D. and Ain, A. and Ajith, P. and Allen, B. and Altin, P. A. and Amariutei, D. V. and Anderson, S. B. and Anderson, W. G. and Arai, K. and Araya, M. C. and Arceneaux, C. C. and Areeda, J. S. and Arun, K. G. and Ashton, G. and Ast, M. and Aston, S. M. and Aufmuth, P. and Aulbert, C. and Babak, S. and Baker, P. T. and Ballmer, S. W. and Barayoga, J. C. and Barclay, S. E. and Barish, B. C. and Barker, D. and Barr, B. and Barsotti, L. and Bartlett, J. and Bartos, I. and Bassiri, R. and Batch, J. C. and Baune, C. and Behnke, B. and Bell, A. S. and Bell, C. J. and Berger, B. K. and Bergman, J. and Bergmann, G. and Berry, C. P. L. and Betzwieser, J. and Bhagwat, S. and Bhandare, R. and Bilenko, I. A. and Billingsley, G. and Birch, J. and Birney, R. and Biscans, S. and Bisht, A. and Biwer, C. and Blackburn, J. K. and Blair, C. D. and Blair, D. and Blair, R. M. and Bock, O. and Bodiya, T. P. and Bogan, C. and Bohe, A. and Bojtos, P. and Bond, C. and Bork, R. and Bose, S. and Brady, P. R. and Braginsky, V. B. and Brau, J. E. and Brinkmann, M. and Brockill, P. and Brooks, A. F. and Brown, D. A. and Brown, D. D. and Brown, N. M. and Buchanan, C. C. and Buikema, A. and Buonanno, A. and Byer, R. L. and Cadonati, L. and Cahillane, C. and Calder\'on Bustillo, J. and Callister, T. and Camp, J. B. and Cannon, K. C. and Cao, J. and Capano, C. D. and Caride, S. and Caudill, S. and Cavagli\`a, M. and Cepeda, C. and Chakraborty, R. and Chalermsongsak, T. and Chamberlin, S. J. and Chan, M. and Chao, S. and Charlton, P. and Chen, H. Y. and Chen, Y. and Cheng, C. and Cho, H. S. and Cho, M. and Chow, J. H. and Christensen, N. and Chu, Q. and Chung, S. and Ciani, G. and Clara, F. and Clark, J. A. and Collette, C. G. and Cominsky, L. and Constancio, M. and Cook, D. and Corbitt, T. R. and Cornish, N. and Corsi, A. and Costa, C. A. and Coughlin, M. W. and Coughlin, S. B. and Countryman, S. T. and Couvares, P. and Coward, D. M. and Cowart, M. J. and Coyne, D. C. and Coyne, R. and Craig, K. and Creighton, J. D. E. and Cripe, J. and Crowder, S. G. and Cumming, A. and Cunningham, L. and Dal Canton, T. and Danilishin, S. L. and Danzmann, K. and Darman, N. S. and Dave, I. and Daveloza, H. P. and Davies, G. S. and Daw, E. J. and DeBra, D. and Del Pozzo, W. and Denker, T. and Dent, T. and Dergachev, V. and DeRosa, R. and DeSalvo, R. and Dhurandhar, S. and D\'{\i}az, M. C. and Di Palma, I. and Dojcinoski, G. and Donovan, F. and Dooley, K. L. and Doravari, S. and Douglas, R. and Downes, T. P. and Drago, M. and Drever, R. W. P. and Driggers, J. C. and Du, Z. and Dwyer, S. E. and Edo, T. B. and Edwards, M. C. and Effler, A. and Eggenstein, H.-B. and Ehrens, P. and Eichholz, J. and Eikenberry, S. S. and Engels, W. and Essick, R. C. and Etzel, T. and Evans, M. and Evans, T. M. and Everett, R. and Factourovich, M. and Fair, H. and Fairhurst, S. and Fan, X. and Fang, Q. and Farr, B. and Farr, W. M. and Favata, M. and Fays, M. and Fehrmann, H. and Fejer, M. M. and Ferreira, E. C. and Fisher, R. P. and Fletcher, M. and Frei, Z. and Freise, A. and Frey, R. and Fricke, T. T. and Fritschel, P. and Frolov, V. V. and Fulda, P. and Fyffe, M. and Gabbard, H. A. G. and Gair, J. R. and Gaonkar, S. G. and Gaur, G. and Gehrels, N. and George, J. and Gergely, L. and Ghosh, A. and Giaime, J. A. and Giardina, K. D. and Gill, K. and Glaefke, A. and Goetz, E. and Goetz, R. and Gondan, L. and Gonz\'alez, G. and Gopakumar, A. and Gordon, N. A. and Gorodetsky, M. L. and Gossan, S. E. and Graef, C. and Graff, P. B. and Grant, A. and Gras, S. and Gray, C. and Green, A. C. and Grote, H. and Grunewald, S. and Guo, X. and Gupta, A. and Gupta, M. K. and Gushwa, K. E. and Gustafson, E. K. and Gustafson, R. and Hacker, J. J. and Hall, B. R. and Hall, E. D. and Hammond, G. and Haney, M. and Hanke, M. M. and Hanks, J. and Hanna, C. and Hannam, M. D. and Hanson, J. and Hardwick, T. and Harry, G. M. and Harry, I. W. and Hart, M. J. and Hartman, M. T. and Haster, C.-J. and Haughian, K. and Heintze, M. C. and Hendry, M. and Heng, I. S. and Hennig, J. and Heptonstall, A. W. and Heurs, M. and Hild, S. and Hoak, D. and Hodge, K. A. and Hollitt, S. E. and Holt, K. and Holz, D. E. and Hopkins, P. and Hosken, D. J. and Hough, J. and Houston, E. A. and Howell, E. J. and Hu, Y. M. and Huang, S. and Huerta, E. A. and Hughey, B. and Husa, S. and Huttner, S. H. and Huynh-Dinh, T. and Idrisy, A. and Indik, N. and Ingram, D. R. and Inta, R. and Isa, H. N. and Isi, M. and Islas, G. and Isogai, T. and Iyer, B. R. and Izumi, K. and Jang, H. and Jani, K. and Jawahar, S. and Jim\'enez-Forteza, F. and Johnson, W. W. and Jones, D. I. and Jones, R. and Ju, L. and K., Haris and Kalaghatgi, C. V. and Kalogera, V. and Kandhasamy, S. and Kang, G. and Kanner, J. B. and Karki, S. and Kasprzack, M. and Katsavounidis, E. and Katzman, W. and Kaufer, S. and Kaur, T. and Kawabe, K. and Kawazoe, F. and Kehl, M. S. and Keitel, D. and Kelley, D. B. and Kells, W. and Kennedy, R. and Key, J. S. and Khalaidovski, A. and Khalili, F. Y. and Khan, S. and Khan, Z. and Khazanov, E. A. and Kijbunchoo, N. and Kim, C. and Kim, J. and Kim, K. and Kim, N. and Kim, N. and Kim, Y.-M. and King, E. J. and King, P. J. and Kinzel, D. L. and Kissel, J. S. and Kleybolte, L. and Klimenko, S. and Koehlenbeck, S. M. and Kokeyama, K. and Kondrashov, V. and Kontos, A. and Korobko, M. and Korth, W. Z. and Kozak, D. B. and Kringel, V. and Krueger, C. and Kuehn, G. and Kumar, P. and Kuo, L. and Lackey, B. D. and Landry, M. and Lange, J. and Lantz, B. and Lasky, P. D. and Lazzarini, A. and Lazzaro, C. and Leaci, P. and Leavey, S. and Lebigot, E. O. and Lee, C. H. and Lee, H. K. and Lee, H. M. and Lee, K. and Lenon, A. and Leong, J. R. and Levin, Y. and Levine, B. M. and Li, T. G. F. and Libson, A. and Littenberg, T. B. and Lockerbie, N. A. and Logue, J. and Lombardi, A. L. and Lord, J. E. and Lormand, M. and Lough, J. D. and L\"uck, H. and Lundgren, A. P. and Luo, J. and Lynch, R. and Ma, Y. and MacDonald, T. and Machenschalk, B. and MacInnis, M. and Macleod, D. M. and Maga\~na-Sandoval, F. and Magee, R. M. and Mageswaran, M. and Mandel, I. and Mandic, V. and Mangano, V. and Mansell, G. L. and Manske, M. and M\'arka, S. and M\'arka, Z. and Markosyan, A. S. and Maros, E. and Martin, I. W. and Martin, R. M. and Martynov, D. V. and Marx, J. N. and Mason, K. and Massinger, T. J. and Masso-Reid, M. and Matichard, F. and Matone, L. and Mavalvala, N. and Mazumder, N. and Mazzolo, G. and McCarthy, R. and McClelland, D. E. and McCormick, S. and McGuire, S. C. and McIntyre, G. and McIver, J. and McManus, D. J. and McWilliams, S. T. and Meadors, G. D. and Melatos, A. and Mendell, G. and Mendoza-Gandara, D. and Mercer, R. A. and Merilh, E. and Meshkov, S. and Messenger, C. and Messick, C. and Meyers, P. M. and Miao, H. and Middleton, H. and Mikhailov, E. E. and Mukund, K. N. and Miller, J. and Millhouse, M. and Ming, J. and Mirshekari, S. and Mishra, C. and Mitra, S. and Mitrofanov, V. P. and Mitselmakher, G. and Mittleman, R. and Mohapatra, S. R. P. and Moore, B. C. and Moore, C. J. and Moraru, D. and Moreno, G. and Morriss, S. R. and Mossavi, K. and Mow-Lowry, C. M. and Mueller, C. L. and Mueller, G. and Muir, A. W. and Mukherjee, Arunava and Mukherjee, D. and Mukherjee, S. and Mullavey, A. and Munch, J. and Murphy, D. J. and Murray, P. G. and Mytidis, A. and Nayak, R. K. and Necula, V. and Nedkova, K. and Neunzert, A. and Newton, G. and Nguyen, T. T. and Nielsen, A. B. and Nitz, A. and Nolting, D. and Normandin, M. E. N. and Nuttall, L. K. and Oberling, J. and Ochsner, E. and O'Dell, J. and Oelker, E. and Ogin, G. H. and Oh, J. J. and Oh, S. H. and Ohme, F. and Oliver, M. and Oppermann, P. and Oram, Richard J. and O'Reilly, B. and O'Shaughnessy, R. and Ott, C. D. and Ottaway, D. J. and Ottens, R. S. and Overmier, H. and Owen, B. J. and Pai, A. and Pai, S. A. and Palamos, J. R. and Palashov, O. and Pal-Singh, A. and Pan, H. and Pankow, C. and Pannarale, F. and Pant, B. C. and Papa, M. A. and Paris, H. R. and Parker, W. and Pascucci, D. and Patrick, Z. and Pearlstone, B. L. and Pedraza, M. and Pekowsky, L. and Pele, A. and Penn, S. and Pereira, R. and Perreca, A. and Phelps, M. and Pierro, V. and Pinto, I. M. and Pitkin, M. and Post, A. and Powell, J. and Prasad, J. and Predoi, V. and Premachandra, S. S. and Prestegard, T. and Price, L. R. and Principe, M. and Privitera, S. and Prokhorov, L. and Puncken, O. and P\"urrer, M. and Qi, H. and Qin, J. and Quetschke, V. and Quintero, E. A. and Quitzow-James, R. and Raab, F. J. and Rabeling, D. S. and Radkins, H. and Raffai, P. and Raja, S. and Rakhmanov, M. and Raymond, V. and Read, J. and Reed, C. M. and Reid, S. and Reitze, D. H. and Rew, H. and Riles, K. and Robertson, N. A. and Robie, R. and Rollins, J. G. and Roma, V. J. and Romanov, G. and Romie, J. H. and Rowan, S. and R\"udiger, A. and Ryan, K. and Sachdev, S. and Sadecki, T. and Sadeghian, L. and Saleem, M. and Salemi, F. and Samajdar, A. and Sammut, L. and Sanchez, E. J. and Sandberg, V. and Sandeen, B. and Sanders, J. R. and Sathyaprakash, B. S. and Saulson, P. R. and Sauter, O. and Savage, R. L. and Sawadsky, A. and Schale, P. and Schilling, R. and Schmidt, J. and Schmidt, P. and Schnabel, R. and Schofield, R. M. S. and Sch\"onbeck, A. and Schreiber, E. and Schuette, D. and Schutz, B. F. and Scott, J. and Scott, S. M. and Sellers, D. and Sergeev, A. and Serna, G. and Sevigny, A. and Shaddock, D. A. and Shahriar, M. S. and Shaltev, M. and Shao, Z. and Shapiro, B. and Shawhan, P. and Sheperd, A. and Shoemaker, D. H. and Shoemaker, D. M. and Siemens, X. and Sigg, D. and Silva, A. D. and Simakov, D. and Singer, A. and Singer, L. P. and Singh, A. and Singh, R. and Sintes, A. M. and Slagmolen, B. J. J. and Smith, J. R. and Smith, N. D. and Smith, R. J. E. and Son, E. J. and Sorazu, B. and Souradeep, T. and Srivastava, A. K. and Staley, A. and Steinke, M. and Steinlechner, J. and Steinlechner, S. and Steinmeyer, D. and Stephens, B. C. and Stone, R. and Strain, K. A. and Strauss, N. A. and Strigin, S. and Sturani, R. and Stuver, A. L. and Summerscales, T. Z. and Sun, L. and Sutton, P. J. and Szczepa\ifmmode \acute{n}\else \'{n}\fi{}czyk, M. J. and Talukder, D. and Tanner, D. B. and T\'apai, M. and Tarabrin, S. P. and Taracchini, A. and Taylor, R. and Theeg, T. and Thirugnanasambandam, M. P. and Thomas, E. G. and Thomas, M. and Thomas, P. and Thorne, K. A. and Thorne, K. S. and Thrane, E. and Tiwari, V. and Tokmakov, K. V. and Tomlinson, C. and Torres, C. V. and Torrie, C. I. and T\"oyr\"a, D. and Traylor, G. and Trifir\`o, D. and Tse, M. and Tuyenbayev, D. and Ugolini, D. and Unnikrishnan, C. S. and Urban, A. L. and Usman, S. A. and Vahlbruch, H. and Vajente, G. and Valdes, G. and Vander-Hyde, D. C. and van Veggel, A. A. and Vass, S. and Vaulin, R. and Vecchio, A. and Veitch, J. and Veitch, P. J. and Venkateswara, K. and Vinciguerra, S. and Vine, D. J. and Vitale, S. and Vo, T. and Vorvick, C. and Vousden, W. D. and Vyatchanin, S. P. and Wade, A. R. and Wade, L. E. and Wade, M. and Walker, M. and Wallace, L. and Walsh, S. and Wang, H. and Wang, M. and Wang, X. and Wang, Y. and Ward, R. L. and Warner, J. and Weaver, B. and Weinert, M. and Weinstein, A. J. and Weiss, R. and Welborn, T. and Wen, L. and We\ss{}els, P. and Westphal, T. and Wette, K. and Whelan, J. T. and White, D. J. and Whiting, B. F. and Williams, R. D. and Williamson, A. R. and Willis, J. L. and Willke, B. and Wimmer, M. H. and Winkler, W. and Wipf, C. C. and Wittel, H. and Woan, G. and Worden, J. and Wright, J. L. and Wu, G. and Yablon, J. and Yam, W. and Yamamoto, H. and Yancey, C. C. and Yap, M. J. and Yu, H. and Zanolin, M. and Zevin, M. and Zhang, F. and Zhang, L. and Zhang, M. and Zhang, Y. and Zhao, C. and Zhou, M. and Zhou, Z. and Zhu, X. J. and Zucker, M. E. and Zuraw, S. E. and Zweizig, J.},
  collaboration = {LIGO Scientific Collaboration},
  journal = {Phys. Rev. D},
  volume = {95},
  issue = {6},
  pages = {062003},
  numpages = {16},
  year = {2017},
  month = {03},
  publisher = {American Physical Society},
  doi = {10.1103/PhysRevD.95.062003},
  url = {https://link.aps.org/doi/10.1103/PhysRevD.95.062003}
}

@article{idq,
doi = {10.1088/2632-2153/abab5f},
url = {https://dx.doi.org/10.1088/2632-2153/abab5f},
year = {2020},
month = {12},
publisher = {IOP Publishing},
volume = {2},
number = {1},
pages = {015004},
author = {Reed Essick and Patrick Godwin and Chad Hanna and Lindy Blackburn and Erik Katsavounidis},
title = {{iDQ: Statistical inference of non-gaussian noise with auxiliary degrees of freedom in gravitational-wave detectors}},
journal = {Machine Learning: Science and Technology}
}

@article{gwskynet,
	doi = {10.3847/2041-8213/abc5b5},
	url = {https://doi.org/10.3847\%2F2041-8213\%2Fabc5b5},
	year = 2020,
	month = {11},
	publisher = {American Astronomical Society},
	volume = {904},
	number = {1},
	pages = {L9},
	author = {Miriam Cabero and Ashish Mahabal and Jess McIver},
	title = {\texttt{GWSkyNet}: A Real-time Classifier for Public Gravitational-wave Candidates},
	journal = {The Astrophysical Journal Letters}
}

@article{o3a,
	title = {{GWTC}-2: {Compact} {Binary} {Coalescences} {Observed} by {LIGO} and {Virgo} {During} the {First} {Half} of the {Third} {Observing} {Run}},
	shorttitle = {{GWTC}-2},
	url = {https://arxiv.org/abs/2010.14527v3},
	doi = {10.1103/PhysRevX.11.021053},
	language = {en},
	urldate = {2022-11-18},
	author = {Abbott, B. P. and Abbott, R. and Abbott, T. D. and Abernathy, M. R. and Acernese, F. and Ackley, K. and Adams, C. and Adams, T. and Addesso, P. and Adhikari, R. X. and Adya, V. B. and Affeldt, C. and Agathos, M. and Agatsuma, K. and Aggarwal, N. and Aguiar, O. D. and Aiello, L. and Ain, A. and Ajith, P. and Allen, B. and Allocca, A. and Altin, P. A. and Anderson, S. B. and Anderson, W. G. and Arai, K. and Arain, M. A. and Araya, M. C. and Arceneaux, C. C. and Areeda, J. S. and Arnaud, N. and Arun, K. G. and Ascenzi, S. and Ashton, G. and Ast, M. and Aston, S. M. and Astone, P. and Aufmuth, P. and Aulbert, C. and Babak, S. and Bacon, P. and Bader, M. K. M. and Baker, P. T. and Baldaccini, F. and Ballardin, G. and Ballmer, S. W. and Barayoga, J. C. and Barclay, S. E. and Barish, B. C. and Barker, D. and Barone, F. and Barr, B. and Barsotti, L. and Barsuglia, M. and Barta, D. and Bartlett, J. and Barton, M. A. and Bartos, I. and Bassiri, R. and Basti, A. and Batch, J. C. and Baune, C. and Bavigadda, V. and Bazzan, M. and Behnke, B. and Bejger, M. and Belczynski, C. and Bell, A. S. and Bell, C. J. and Berger, B. K. and Bergman, J. and Bergmann, G. and Berry, C. P. L. and Bersanetti, D. and Bertolini, A. and Betzwieser, J. and Bhagwat, S. and Bhandare, R. and Bilenko, I. A. and Billingsley, G. and Birch, J. and Birney, R. and Birnholtz, O. and Biscans, S. and Bisht, A. and Bitossi, M. and Biwer, C. and Bizouard, M. A. and Blackburn, J. K. and Blair, C. D. and Blair, D. G. and Blair, R. M. and Bloemen, S. and Bock, O. and Bodiya, T. P. and Boer, M. and Bogaert, G. and Bogan, C. and Bohe, A. and Bojtos, P. and Bond, C. and Bondu, F. and Bonnand, R. and Boom, B. A. and Bork, R. and Boschi, V. and Bose, S. and Bouffanais, Y. and Bozzi, A. and Bradaschia, C. and Brady, P. R. and Braginsky, V. B. and Branchesi, M. and Brau, J. E. and Briant, T. and Brillet, A. and Brinkmann, M. and Brisson, V. and Brockill, P. and Brooks, A. F. and Brown, D. A. and Brown, D. D. and Brown, N. M. and Buchanan, C. C. and Buikema, A. and Bulik, T. and Bulten, H. J. and Buonanno, A. and Buskulic, D. and Buy, C. and Byer, R. L. and Cabero, M. and Cadonati, L. and Cagnoli, G. and Cahillane, C. and Bustillo, J. Calder\'on and Callister, T. and Calloni, E. and Camp, J. B. and Cannon, K. C. and Cao, J. and Capano, C. D. and Capocasa, E. and Carbognani, F. and Caride, S. and Diaz, J. Casanueva and Casentini, C. and Caudill, S. and Cavagli\`a, M. and Cavalier, F. and Cavalieri, R. and Cella, G. and Cepeda, C. B. and Baiardi, L. Cerboni and Cerretani, G. and Cesarini, E. and Chakraborty, R. and Chalermsongsak, T. and Chamberlin, S. J. and Chan, M. and Chao, S. and Charlton, P. and Chassande-Mottin, E. and Chen, H. Y. and Chen, Y. and Cheng, C. and Chincarini, A. and Chiummo, A. and Cho, H. S. and Cho, M. and Chow, J. H. and Christensen, N. and Chu, Q. and Chua, S. and Chung, S. and Ciani, G. and Clara, F. and Clark, J. A. and Cleva, F. and Coccia, E. and Cohadon, P.-F. and Colla, A. and Collette, C. G. and Cominsky, L. and Constancio, M. and Conte, A. and Conti, L. and Cook, D. and Corbitt, T. R. and Cornish, N. and Corsi, A. and Cortese, S. and Costa, C. A. and Coughlin, M. W. and Coughlin, S. B. and Coulon, J.-P. and Countryman, S. T. and Couvares, P. and Cowan, E. E. and Coward, D. M. and Cowart, M. J. and Coyne, D. C. and Coyne, R. and Craig, K. and Creighton, J. D. E. and Creighton, T. D. and Cripe, J. and Crowder, S. G. and Cruise, A. M. and Cumming, A. and Cunningham, L. and Cuoco, E. and Canton, T. Dal and Danilishin, S. L. and D'Antonio, S. and Danzmann, K. and Darman, N. S. and Da Silva Costa, C. F. and Dattilo, V. and Dave, I. and Daveloza, H. P. and Davier, M. and Davies, G. S. and Daw, E. J. and Day, R. and De, S. and DeBra, D. and Debreczeni, G. and Degallaix, J. and De Laurentis, M. and Del\'eglise, S. and Del Pozzo, W. and Denker, T. and Dent, T. and Dereli, H. and Dergachev, V. and DeRosa, R. T. and De Rosa, R. and DeSalvo, R. and Dhurandhar, S. and D\'{\i}az, M. C. and Di Fiore, L. and Di Giovanni, M. and Di Lieto, A. and Di Pace, S. and Di Palma, I. and Di Virgilio, A. and Dojcinoski, G. and Dolique, V. and Donovan, F. and Dooley, K. L. and Doravari, S. and Douglas, R. and Downes, T. P. and Drago, M. and Drever, R. W. P. and Driggers, J. C. and Du, Z. and Ducrot, M. and Dwyer, S. E. and Edo, T. B. and Edwards, M. C. and Effler, A. and Eggenstein, H.-B. and Ehrens, P. and Eichholz, J. and Eikenberry, S. S. and Engels, W. and Essick, R. C. and Etzel, T. and Evans, M. and Evans, T. M. and Everett, R. and Factourovich, M. and Fafone, V. and Fair, H. and Fairhurst, S. and Fan, X. and Fang, Q. and Farinon, S. and Farr, B. and Farr, W. M. and Favata, M. and Fays, M. and Fehrmann, H. and Fejer, M. M. and Feldbaum, D. and Ferrante, I. and Ferreira, E. C. and Ferrini, F. and Fidecaro, F. and Finn, L. S. and Fiori, I. and Fiorucci, D. and Fisher, R. P. and Flaminio, R. and Fletcher, M. and Fong, H. and Fournier, J.-D. and Franco, S. and Frasca, S. and Frasconi, F. and Frede, M. and Frei, Z. and Freise, A. and Frey, R. and Frey, V. and Fricke, T. T. and Fritschel, P. and Frolov, V. V. and Fulda, P. and Fyffe, M. and Gabbard, H. A. G. and Gair, J. R. and Gammaitoni, L. and Gaonkar, S. G. and Garufi, F. and Gatto, A. and Gaur, G. and Gehrels, N. and Gemme, G. and Gendre, B. and Genin, E. and Gennai, A. and George, J. and Gergely, L. and Germain, V. and Ghosh, Abhirup and Ghosh, Archisman and Ghosh, S. and Giaime, J. A. and Giardina, K. D. and Giazotto, A. and Gill, K. and Glaefke, A. and Gleason, J. R. and Goetz, E. and Goetz, R. and Gondan, L. and Gonz\'alez, G. and Castro, J. M. Gonzalez and Gopakumar, A. and Gordon, N. A. and Gorodetsky, M. L. and Gossan, S. E. and Gosselin, M. and Gouaty, R. and Graef, C. and Graff, P. B. and Granata, M. and Grant, A. and Gras, S. and Gray, C. and Greco, G. and Green, A. C. and Greenhalgh, R. J. S. and Groot, P. and Grote, H. and Grunewald, S. and Guidi, G. M. and Guo, X. and Gupta, A. and Gupta, M. K. and Gushwa, K. E. and Gustafson, E. K. and Gustafson, R. and Hacker, J. J. and Hall, B. R. and Hall, E. D. and Hammond, G. and Haney, M. and Hanke, M. M. and Hanks, J. and Hanna, C. and Hannam, M. D. and Hanson, J. and Hardwick, T. and Harms, J. and Harry, G. M. and Harry, I. W. and Hart, M. J. and Hartman, M. T. and Haster, C.-J. and Haughian, K. and Healy, J. and Heefner, J. and Heidmann, A. and Heintze, M. C. and Heinzel, G. and Heitmann, H. and Hello, P. and Hemming, G. and Hendry, M. and Heng, I. S. and Hennig, J. and Heptonstall, A. W. and Heurs, M. and Hild, S. and Hoak, D. and Hodge, K. A. and Hofman, D. and Hollitt, S. E. and Holt, K. and Holz, D. E. and Hopkins, P. and Hosken, D. J. and Hough, J. and Houston, E. A. and Howell, E. J. and Hu, Y. M. and Huang, S. and Huerta, E. A. and Huet, D. and Hughey, B. and Husa, S. and Huttner, S. H. and Huynh-Dinh, T. and Idrisy, A. and Indik, N. and Ingram, D. R. and Inta, R. and Isa, H. N. and Isac, J.-M. and Isi, M. and Islas, G. and Isogai, T. and Iyer, B. R. and Izumi, K. and Jacobson, M. B. and Jacqmin, T. and Jang, H. and Jani, K. and Jaranowski, P. and Jawahar, S. and Jim\'enez-Forteza, F. and Johnson, W. W. and Johnson-McDaniel, N. K. and Jones, D. I. and Jones, R. and Jonker, R. J. G. and Ju, L. and Haris, K. and Kalaghatgi, C. V. and Kalogera, V. and Kandhasamy, S. and Kang, G. and Kanner, J. B. and Karki, S. and Kasprzack, M. and Katsavounidis, E. and Katzman, W. and Kaufer, S. and Kaur, T. and Kawabe, K. and Kawazoe, F. and K\'ef\'elian, F. and Kehl, M. S. and Keitel, D. and Kelley, D. B. and Kells, W. and Kennedy, R. and Keppel, D. G. and Key, J. S. and Khalaidovski, A. and Khalili, F. Y. and Khan, I. and Khan, S. and Khan, Z. and Khazanov, E. A. and Kijbunchoo, N. and Kim, C. and Kim, J. and Kim, K. and Kim, Nam-Gyu and Kim, Namjun and Kim, Y.-M. and King, E. J. and King, P. J. and Kinzel, D. L. and Kissel, J. S. and Kleybolte, L. and Klimenko, S. and Koehlenbeck, S. M. and Kokeyama, K. and Koley, S. and Kondrashov, V. and Kontos, A. and Koranda, S. and Korobko, M. and Korth, W. Z. and Kowalska, I. and Kozak, D. B. and Kringel, V. and Krishnan, B. and Kr\'olak, A. and Krueger, C. and Kuehn, G. and Kumar, P. and Kumar, R. and Kuo, L. and Kutynia, A.},
  collaboration = {LIGO Scientific Collaboration and Virgo Collaboration},
  }

@article{powell,
author = {Powell, Jade and Torres-Forné, Alejandro and Lynch, Roseann and Trifirò, Daniele and Cuoco, Elena and Cavaglià, Marco and Heng, Ik and Font, Jose},
year = {2017},
month = {02},
pages = {},
title = {{Classification methods for noise transients in advanced gravitational-wave detectors II: Performance tests on Advanced LIGO data}},
volume = {34},
journal = {Classical and Quantum Gravity},
doi = {10.1088/1361-6382/34/3/034002}
}

@article{gwosc,
title = {{GWTC-3 Data Release}},
year = {2021},
doi = {https://www.gw-openscience.org/GWTC-3/},
url = {https://www.gw-openscience.org/GWTC-3/},
author = {{LIGO Scientific Collaboration, Virgo Collaboration
and KAGRA Collaboration}}
}

@article{gracedb,
author={{Pace A, Prestegard T, Moe B and Stephens B}},
title = {{GraceDB—Gravitational-Wave
Candidate Event Database}},
year = {2020},
doi = {https://gracedb.ligo.org/},
url = {https://gracedb.ligo.org/}
}

@article{dqr,
author={{The LIGO Scientific Collaboration and The Virgo Collaboration}},
title = {{Data Quality Report user
documentation}},
year = {2018},
doi = {https://docs.ligo.org/detchar/data-quality-report/},
url = {https://docs.ligo.org/detchar/data-quality-report/}
}

@book{geron,
  title={{Hands-on machine learning with Scikit-Learn, Keras, and TensorFlow}},
  author={G{\'e}ron, Aur{\'e}lien},
  year={2022},
  publisher={{O'Reilly Media, Inc.}}
}

@article{gspyO3,
	doi = {10.1088/1361-6382/ac1ccb},
	url = {https://doi.org/10.1088\%2F1361-6382\%2Fac1ccb},
	year = 2021,
	month = {09},
	publisher = {{IOP} Publishing},
	volume = {38},
	number = {19},
	pages = {195016},
	author = {S Soni and C P L Berry and S B Coughlin and M Harandi and C B Jackson and K Crowston and C {\O}sterlund and O Patane and A K Katsaggelos and L Trouille and V-G Baranowski and W F Domainko and K Kaminski and M A Lobato Rodriguez and U Marciniak and P Nauta and G Niklasch and R R Rote and B T{\'{e}
}gl{\'{a}}s and C Unsworth and C Zhang},
	title = {Discovering features in gravitational-wave data through detector characterization, citizen science, and machine learning},
	journal = {Classical and Quantum Gravity}
}

@article{derek_validation,
	doi = {10.1088/1361-6382/aca238},
	url = {https://doi.org/10.1088\%2F1361-6382\%2Faca238},
	year = 2022,
	month = {11},
	publisher = {{IOP} Publishing},
	volume = {39},
	number = {24},
	pages = {245013},
	author = {D Davis and T B Littenberg and I M Romero-Shaw and M Millhouse and J McIver and F Di Renzo and G Ashton},
	title = {Subtracting glitches from gravitational-wave detector data during the third {LIGO}-Virgo observing run},
	journal = {Classical and Quantum Gravity}
}

@article{macas,
  title = {Impact of noise transients on low latency gravitational-wave event localization},
  author = {Macas, Ronaldas and Pooley, Joshua and Nuttall, Laura K. and Davis, Derek and Dyer, Martin J. and Lecoeuche, Yannick and Lyman, Joseph D. and McIver, Jess and Rink, Katherine},
  journal = {Phys. Rev. D},
  volume = {105},
  issue = {10},
  pages = {103021},
  numpages = {16},
  year = {2022},
  month = {05},
  publisher = {American Physical Society},
  doi = {10.1103/PhysRevD.105.103021},
  url = {https://link.aps.org/doi/10.1103/PhysRevD.105.103021}
}

\end{document}